\documentclass[preprint]{aastex62}

\usepackage{acronym}
\usepackage{booktabs}
\usepackage{dcolumn}
\usepackage{graphicx}
\usepackage{amsmath,amsfonts,amssymb}
\usepackage{mathrsfs}
\usepackage[normalem]{ulem}
\graphicspath{{./}{figures/}}

\shorttitle{Magnification Statistics with Microlensing}
\shortauthors{Dai et. al.}

\newcommand{\refeq}[1]{Eq.~(\ref{eq:#1})}          
          
\newcommand{\reffig}[1]{Figure~\ref{fig:#1}}          
\newcommand{\refsec}[1]{Section~\ref{sec:#1}}
\newcommand{\refapp}[1]{Appendix~\ref{app:#1}}

\newcommand{\be}{\begin{equation}}
\newcommand{\ee}{\end{equation}}
\newcommand{\ba}{\begin{eqnarray}}
\newcommand{\ea}{\end{eqnarray}}

\newcommand{\bfzero}{\boldsymbol{0}}
\newcommand{\bfx}{\boldsymbol{x}}
\newcommand{\bfy}{\boldsymbol{y}}
\newcommand{\bfz}{\boldsymbol{z}}

\newcommand{\bfalp}{\boldsymbol{\alpha}}
\newcommand{\bfell}{\boldsymbol{\ell}}
\newcommand{\bfd}{\boldsymbol{d}}
\newcommand{\bfr}{\boldsymbol{r}}

\newcommand{\bfl}{{\boldsymbol{\ell}}}

\newcommand{\bfR}{{\boldsymbol{R}}}
\newcommand{\bfa}{\boldsymbol{a}}
\newcommand{\bfb}{\boldsymbol{b}}

\newcommand{\Msun}{{\rm M}_\odot}

\def\VEV#1{\left\langle #1 \right\rangle}
\def\rmd{\mathrm{d}}

\begin{document}

\title{New Approximation of Magnification Statistics for Random Microlensing of Magnified Sources}

\correspondingauthor{Liang Dai}
\email{liangdai@berkeley.edu}

\author[0000-0003-2091-8946]{Liang Dai}
\affiliation{Department of Physics, University of California, 366 Physics North MC 7300, Berkeley, CA 94720, USA}

\author{Massimo Pascale}
\affiliation{Department of Astronomy, University of California, 501 Campbell Hall \#3411, Berkeley, CA 94720, USA}



\begin{abstract}

Gravitationally lensed extragalactic sources are often subject to statistical microlensing by stars in the galaxy or cluster lens. Accurate models of the flux statistics are required for inferring source and lens properties from flux observations. We derive an accurate semi-analytic approximation for calculating the mean and variance of the magnification factor, which are applicable to Gaussian source profiles and arbitrary non-uniform macro lens models, and hence can save the need to perform expensive numerical simulations. The results are given as single and double lens-plane integrals with simple, non-oscillatory integrands, and hence are fast computable using common Monte Carlo integrators. Employing numerical ray-shooting experiments, we examine the case of a highly magnified source near a macro fold caustic, and demonstrate the excellent accuracy of this semi-analytic approximation in the regime of multiple micro images. Additionally, we point out how the maximum persistent magnification achievable near a macro caustic is fundamentally limited by the masses and number density of the foreground microlenses, in addition to the source's physical size.

\end{abstract}

\bigskip
$~$

\section{Introduction}
\label{sec:intro}

Strong gravitational lensing of extragalactic sources by intervening galaxies or galaxy clusters are exquisite astrophysical probes. The observed fluxes of these sources are often strongly affected by random microlensing, collectively caused by many compact masses embedded within the foreground lens~\citep{KayserRefsdalStabell1986microlensing, Wambsganss1992muPDFML}. Observations of multiply imaged quasars first engendered theoretical interests in statistical microlensing (see \cite{Wambsganss2006MicrolensingReview} for a review), with the major aim to resolve the size and structure of the quasar accretion disk~\citep{BlackBurne2011QSOmicrolensing} and to probe compact lenses of planetary to stellar masses.

To quantify the flux variability under random microlensing, a statistical approach is appropriate. For observations carried out at random epochs, the key observables include the mean and variance of the magnification factor. The scenario of microlenses embedded in constant background convergence and shear has been extensively studied, mostly in the context of multiply-imaged quasars. In this case, the magnification factor averaged over random microlens realizations equals the macro one regardless of the source size, while the variance is difficult to compute analytically and was the focus of many past works. Lenses solely comprised of compact objects were studied in the pioneering works of \cite{DeguchiWatson1987muvariance} and \cite{RefsdalStabell1991MLLargeSources}, and the effects of a shear and a diffuse surface mass component were investigated in follow-up studies~\citep{SeitzSchneider1994microlensingI, Seitz1994microlensingII, RefsdalStabell1997MLLargeSrcShear}. The statistical formalism was further developed in more recent works~\citep{Neindorf2003extragalML, Tuntsov2004compactDMFromClusters, GoodmanSun2014MicrolensingFluxVariance}, and other methods have been explored~\citep{Fleury2020AnalyticAmpStats}. When analytical results are unavailable, unreliable, or cumbersome to apply, direct numerical simulations come to rescue. Important numerical techniques that greatly enhance efficiency and accuracy include inverse ray-shooting~\citep{KayserRefsdalStabell1986microlensing}, the hierarchical tree algorithm~\citep{Wambsganss1999MLTreeAlgorithm}, and the image tracking method~\citep{Lewis1993imagetrack, Witt1993imagetrack}.

Recently, theoretical interests in statistical microlensing have been reinvigorated by the detections of high-$z$ stars~\citep{1991ApJ...379...94M} magnified by a spectacular $\sim 10^2$--$10^3$ folds near critical curves of cluster lenses~\citep{2018NatAs...2..334K, 2018NatAs...2..324R, Chen:2019ncy, 2019ApJ...880...58K}. Owing to the extreme macro magnification factors realized in these cases, even a low convergence of intracluster stars around the cluster Einstein radius $\kappa_\star\sim 10^{-3}$--$10^{-2}$ (compared to $\kappa_\star \sim 0.1$--$1$ around galaxy Einstein radii) leads to frequent microlensing brightening episodes~\citep{2017ApJ...850...49V, 2018ApJ...857...25D, Oguri:2017ock, Diego2019ExtremeMagnificationUniverse}, making these outstanding probes of compact constituents of the lens mass. Such stochastic microlensing may act also on a cluster of stars if the cluster as a whole is highly magnified~\citep{Dai2020ArcSymmetryS1226, Dai2021SunburstStarClusterMicrolensing}. With new observational data, other highly magnified candidate sources: lensed quasar images suspected of large magnification~\citep{Fujimoto2020UltraluminousQuasar} or flux anomaly~\citep{Glikman2018arXiv180705434G} have been reported; a magnified ``knot'' in a Cosmic Noon starburst showing perplexing flux anomalies~\citep{Vanzella2020SunburstTr}. 

While semi-analytic scaling laws have offered much insight into the behavior of statistical microlensing in the extreme magnification regime, more accurate modeling of flux statistics have so far relied on large yet artful numerical simulations~\citep{2017ApJ...850...49V, 2018ApJ...857...25D, Diego2019ExtremeMagnificationUniverse, Dai2021SunburstStarClusterMicrolensing}. Sometimes, it may even be computationally prohibitive to simulate for realistic parameter values! In this work, with a focus on the macro caustic vicinity, we aim to develop a general and practical analytic model, which will both deepen our understanding of microlensing in the high optical depth regime and facilitate scans of large parameter space.

Microlensing has a significant impact on the magnifications achievable near a macro caustic~\citep{2017ApJ...850...49V, 2018ApJ...857...25D, Diego2019ExtremeMagnificationUniverse}. Microlenses of a characteristic Einstein radius $\theta_\star$ induce a stochastic deflection component $ \kappa^{1/2}_\star\,\theta_\star$ in the ray equation~\citep{Katz1986RandomScattering}, up to a multiplicative ``Coulomb'' logarithm reflecting the long-ranged nature of point lens deflections. This sets an effective smoothing scale on the source plane, independently of the source's angular extent $\sigma_{\rm W}$. If the magnification caused by the macro lens is uniform across this scale, stochastic microlensing conserves the mean magnification, but only induces fluctuations around it. However, when the macro magnification varies significantly, a situation that inevitably arises near a caustic, even the mean magnification is modified. Indeed, \cite{2017ApJ...850...49V} showed that a sharp caustic induced by a macro lens of a smooth mass profile is ``smeared out'' across a width of $\sim \kappa^{1/2}_\star\,\theta_\star$ by microlenses.

In the absence of microlenses, the maximum magnification is realized when a finite source grazes a sharp marco caustic, $\mu_{\rm max} \sim (\sigma_{\rm W}\,d)^{-1/2}$, where $d$ is typically the inverse of the characteristic angular scale of the macro lens that produces the caustic. Sub-galactic substructure lenses produce secondary caustics~\citep{2018ApJ...867...24D} that have larger values of $d$ compared to what smooth galaxy-scale or cluster-scale lenses can produce, and hence decreased values of $\mu_{\rm max}$. In the presence of microlenses, the sharp macro caustic is disrupted, and a corrugated network of micro caustics form instead. In this situation, the random deflection scale $\kappa^{1/2}_\star\,\theta_\star$ comes into play. Now the mean magnification can reach a maximum value $\sim (\sigma_{\rm eff}\,d)^{-1/2}$ where $\sigma_{\rm eff} \simeq \sqrt{\sigma^2_{\rm W} + \kappa_\star\,\theta^2_\star}$, i.e. the grazing scale can get as small as $\sigma_{\rm W}$ or $\kappa^{1/2}_\star\,\theta_\star$, whichever is larger. 

In real astrophysical contexts, a source may be sufficiently large (e.g. quasar, SN) to overlap multiple micro caustics, i.e. $\sigma_{\rm W} \gtrsim \theta_\star\,\kappa^{-1/2}_\star/(\sigma_{\rm eff}\,d)^{-1/2}$, when the density of micro caustics is the highest in the proximity of a macro caustic. As a result, its flux fluctuations around the mean do not exceed the mean by any large factor, and hence $(\sigma_{\rm eff}\,d)^{-1/2}$ is often a fair estimate for the maximal possible magnification after accounting for fluctuations. For smaller sources, it may still be true that $\sigma_{\rm eff} \gtrsim \theta_\star\,\kappa^{-1/2}_\star/(\sigma_{\rm eff}\,d)^{-1/2}$, so that flux fluctuations remain mild as many disconnected micro images form and contribute uncorrelated magnification fluctuations. Even in this regime, $(\sigma_{\rm eff}\,d)^{-1/2}$ is not an underestimate of the maximal achievable magnification. Flux fluctuations can greatly exceed the mean value only for very small sources (e.g. individual stellar photospheres), and for a sufficiently low surface number density of microlenses. During these short events of micro caustic crossing, the peak magnification is dominated by just a pair of micro images.


A conceptual obstacle to fully analytic calculations of the microlensing flux statistics has to do with the ``ultraviolet'' (UV) and ``infrared'' (IR) divergences, which are terms borrowed from field theory. These logarithmic divergences arise because the deflection due to any single point microlens scales inversely with the impact parameter, analogous to the Coulomb divergence associated with the inverse-square force law in three-dimensonal space. The UV divergence is traced back to arbitrarily large ray deflections at arbitrarily small impact parameters to any single point microlens; consequently, the probability distribution functions (PDFs) for the random deflections have divergent higher-order moments, and the corresponding characteristic functions (CFs) contain non-analytic logarithmic terms. We seek a prescription in which the UV divergence is regulated by the finite source size. On the other hand, the IR divergence implies that the statistics can be sensitive to far-away microlenses distributed over the largest scales on the lens plane. We require a prescription in which the dependence on the IR cutoff on statistical averages is manifest and unambiguous.

The main result of this work is an accurate semi-analytic approximation for the mean and variance of the magnifications (\refeq{muWvevML} and \refeq{muW2dblintegML} respectively), for a Gaussian source profile, and in the regime that multiple micro images form. The gist of the approximation is about capturing the Gaussian bulk of the deflection distribution, while judiciously neglecting the non-Gaussian tail of large deflections which are only important for either a small portion of the source or a minority of the micro images. The approximation determines the UV and IR logarithms in a physical way, and respect the expected translational and rotational symmetries of the deflection distributions. The results only require evaluating two- and four-dimensional integrals with simple, well-behaved integrands, and are applicable to {\it any} macro lens model. In practice, answers can be obtained in less than a second by employing standard Monte Carlo integrators. The new results can be directly used to efficiently and accurately quantify the magnification statistics, for microlenses embeded in a variety of macro lens models and as a function of microlens abundance and source size, thus saving the need to perform expensive yet tricky numerical simulations. Conversely, the semi-analytic answers enable calibration to numerical codes.

The remainder of the paper is organized as follows. In \refsec{theory}, we introduce the general theoretical framework to compute magnification statistics, in particular, the mean and the variance. In \refsec{gaussian}, we examine a toy model in which microlensing deflections are Gaussian random variables. We will develop useful intuition into the problem as this toy model is exactly solvable. In \refsec{microlens}, we turn to the real problem of discrete microlenses, and derive our key results. We then demonstrate that our semi-analytic approximation agrees well with numerical ray-shooting experiments, for various parameter choices ranging from Gaussian to non-Gaussian flux variability behaviors. We will discuss our results in \refsec{discuss}, before we give concluding remarks in \refsec{concl}. Additional technical details are presented in Appendices for reference. Results of \refsec{gaussian} and \refsec{microlens} are presented in dimensionless angular units, and can be easily scaled to the appropriate physical units in any specified astrophysical context.

\section{Theoretical framework}
\label{sec:theory}

A dozen of previous studies have treated microlensing using a statistical theory. Those include the earlier works of \cite{DeguchiWatson1987muvariance}, \cite{Katz1986RandomScattering}, \cite{SeitzSchneider1994microlensingI} and \cite{Seitz1994microlensingII}, as well as a more general formulation in \cite{Neindorf2003extragalML}. For clarity, we reintroduce this theoretical framework, which is based on the concept of multivariate probability distribution functions and the corresponding characteristic functions.

We decompose the deflection field into a background component $\bfalp_{\rm B}(\bfx)$ and a fluctuating component $\bfalp(\bfx)$. The former varies smoothly as a function of the image plane position $\bfx$ and is given. The latter, being stochastic in nature with specific spatial correlations, will be given a statistical treatment. We adopt the general assumption that the $\bfalp(\bfx)$'s have correlation functions that are invariant under spatial translations and rotations, as is the case if $\bfalp(\bfx)$ is generated by point-like microlenses that are uniformly distributed on the lens plane. 

For a point source at the source-plane position $\bfy$, the lens equation is $\bfy = \bfx - \bfalp_{\rm B}(\bfx) - \bfalp(\bfx)$. Let $W(\bfy)$ be the normalized surface brightness profile of a finite-sized source centered at the source-plane origin, satisfying $\int\,\rmd^2\bfy\,W(\bfy) = 1$. If unresolved, the total magnification summed over all geometric images can be written as
\begin{align}
    \mu_{\rm W}(\bfy) = \int\,\rmd^2\bfx\,W\left( \bfx - \bfy - \bfalp_{\rm B}(\bfx) - \bfalp(\bfx) \right).
\end{align}
Throughout, we use the $\VEV{\cdots}$ notation to indicate averaging over random realizations of $\bfalp(\bfx)$. Inserting the Fourier decomposition of $W(\bfy)$, we obtain the mean magnification factor~\citep{2017ApJ...850...49V}
\begin{eqnarray}
\label{eq:muWy}
    \VEV{\mu_{\rm W}(\bfy)} & = &  \int\,\rmd^2\bfx\, \int\,\frac{\rmd^2\bfl}{(2\pi)^2}\,e^{-i\,\bfl\cdot(\bfx - \bfy - \bfalp_{\rm B}(\bfx))}\,\widetilde W(\bfl)\,\VEV{e^{i\,\bfl\cdot\bfalp(\bfx)}},
\end{eqnarray}
where $\bfell$ is the Fourier wave vector conjugate to the real-space angular variable, and $\widetilde W(\bfl)= \int\,\rmd^2\bfy\,W(\bfy)\,e^{i\,\bfl\cdot\bfy}$ is the Fourier transform of $W(\bfy)$. This expression depends on the characteristic function (CF) involving the stochastic deflection at one image-plane position $\bfx$. By statistical homogeneity, $\VEV{e^{i\,\bfell\cdot\bfalp(\bfx)}}$ is independent of $\bfx$. The result can be recast into the form
\begin{align}
\label{eq:muWmean}
    \VEV{\mu_{\rm W}(\bfy)} = \int\,\rmd^2\bfx\,P^{\rm W}_1\left( \bfx - \bfy - \bfalp_{\rm B}(\bfx) \right),
\end{align}
where $P^{\rm W}_1(\bfalp)$ is the one-point probability distribution function (PDF) for the random deflection $\bfalp(\bfx)$ at any $\bfx$, $P_1(\bfalp)$ (see e.g. \cite{Katz1986RandomScattering}), convoluted with the source profile
\begin{align}
\label{eq:P1Wdef}
    P^{\rm W}_1(\bfalp) = \int\,\rmd^2\bfy\,W(\bfy)\,P_1(\bfalp - \bfy).
\end{align}
By transforming the integration variable from $\bfy$ to $\bfalp=\bfx - \bfy - \bfalp_{\rm B}(\bfx)$, \refeq{muWmean} can also be written as (see e.g. \cite{2017ApJ...850...49V})
\begin{align}
\label{eq:muWmeanbfalp}
    \VEV{\mu_{\rm W}(\bfy)} = \int\,\rmd^2\bfalp\,P^{\rm W}_1(\bfalp)\,\mu_{\rm B}(\bfy + \bfalp),
\end{align}
where $\mu_{\rm B}(\bfy)$ is the magnification of a point source at $\bfy$ due to only the background deflection. 

\refeq{muWmeanbfalp} can be interpreted as smoothing the point-source background magnification pattern with a ``point spread function'' $P^{\rm W}_1(\bfalp)$, whose characteristic width is set by that of the source profile $W(\bfy)$ or that of the random deflections $P_1(\bfalp)$, whichever is larger. $\VEV{\mu_{\rm W}(\bfy)}$ equals the background magnification $\mu_{\rm B}(\bfy)$ if the latter is approximately uniform over the smoothing scale set by $P^{\rm W}_1(\bfalp)$. A particularly interesting situation arises near a lensing caustic where $\mu_{\rm B}(\bfy)$ often varies rapidly; as a result $\VEV{\mu_{\rm W}(\bfy)}$ can differ substantially from $\mu_{\rm B}(\bfy)$.

The simplest way to quantify the scatter in the value of $\mu_{\rm W}(\bfy)$ due to random microlens realizations is the second moment, which is given by~\citep{Neindorf2003extragalML}
\begin{eqnarray}
\label{eq:muWy2}
    \VEV{\mu_{\rm W}(\bfy)^2} & = &  \int\,\rmd^2\bfx_1\, \int\,\rmd^2\bfx_2\,\int\,\frac{\rmd^2\bfl_1}{(2\pi)^2}\,\int\,\frac{\rmd^2\bfl_2}{(2\pi)^2}\,e^{-i\,\bfl_1\cdot(\bfx_1 - \bfy - \bfalp_{\rm B}(\bfx_1))}\,e^{-i\,\bfl_2\cdot(\bfx_2 - \bfy - \bfalp_{\rm B}(\bfx_2))} \nonumber\\
&& \times \widetilde  W(\bfl_1)\,\widetilde  W(\bfl_2)\,\VEV{e^{i\,\bfl_1\cdot\bfalp(\bfx_1)}\,e^{i\,\bfl_2\cdot\bfalp(\bfx_2)}}.
\end{eqnarray}
The CF involving the stochastic deflections at {\it two} different image-plane positions $\bfx_1$ and $\bfx_2$ needs to be computed. By statistical homogeneity, $\VEV{e^{i\,\bfl_1\cdot\bfalp(\bfx_1)}\,e^{i\,\bfl_2\cdot\bfalp(\bfx_2)}}$ only depends on $\bfx_2 - \bfx_1$, but not on $\bfx_1$ and $\bfx_2$ individually. The magnification factor has a standard deviation ${\rm Std}[\mu_{\rm W}(\bfy)] = \sqrt{\VEV{\mu_{\rm W}(\bfy)^2} - \VEV{\mu_{\rm W}(\bfy)}^2}$~\citep{Neindorf2003extragalML}.

\section{Toy model: Gaussian random deflections}
\label{sec:gaussian}

Before we consider discrete, point microlenses as realistic random deflectors, we would like to first solve a toy model in which $\bfalp(\bfx)$ behaves strictly as a Gaussian random vector field on the image plane. The toy model allows for the analytic calculation of many results, and hence will offer us much insight into how the source-convoluted magnification factor behaves in the presence of stochastic deflections. We note that this Gaussian deflection model precisely describes the collective lensing effects of axion minihalos on lensed extragalatic stars crossing micro caustics~\citep{Dai2020AxionMinihalo}.

For full analytic tractability, we choose to consider a Gaussian source profile with half width $\sigma_{\rm W}$, $W(\bfy) = [(2\pi)\,\sigma^2_{\rm W}]^{-1}\,\exp(-(1/2)\,|\bfy|^2/\sigma^2_{\rm W})$ throughout. While many commonly studied sources such as individual stars are more appropriately modeled as uniform disks, the assumption of a Gaussian source profile does not impose a fundamental limitation of our derivation, as we will see later. The Fourier transform of this Gaussian profile is given by
\begin{align}
\label{eq:Wlgaussian}
\widetilde W(\bfl) = \exp\left(-\frac12\,\sigma^2_{\rm W}\,|\bfl|^2 \right).
\end{align}

We can generally assume that $\bfalp(\bfx)$ is the gradient of a scalar potential (valid for deflection by a single lens plane), $\bfalp(\bfx) = \nabla\,\psi(\bfx)$, and that in the Fourier domain the scalar potential has an isotropic power spectrum $\langle\widetilde\psi(\bfl)\,\widetilde\psi^*(\bfl')\rangle = (2\pi)^2\,\delta_D(\bfl - \bfl')\,P_\psi(\ell)$. The two-point correlation for $\bfalp(\bfx)$ can be decomposed into a longitudinal component $C_\parallel(r)$ and a transverse component $C_\perp(r)$:
\begin{align}
\label{eq:2ptalpalp}
    \VEV{\alpha_i(\bfx_1)\,\alpha_j(\bfx_2)} =  C_\parallel(r
    _{12})\,\frac{r_{12,i}\,r_{12,j}}{r^2_{12}} + C_\perp(r_{12})\,\left( \delta_{ij} - \frac{r_{12,i}\,r_{12,j}}{r^2_{12}} \right),
\end{align}
where $\bfr_{12} = \bfx_2 - \bfx_1$ and $r_{12} = |\bfr_{12}|$. These are related to the potential power spectrum through:
\begin{eqnarray}
\label{eq:Caapara}
    C_\parallel(r) & = & \int\,\frac{\ell^3\,\rmd\ell}{2\pi}\,\left( \frac{J_1(\ell\,r)}{\ell\,r} - J_2(\ell\,r) \right)\,P_\psi(\ell), \\
\label{eq:Caaperp}
C_\perp(r) & = & \int\,\frac{\ell^3\,\rmd\ell}{2\pi}\, \frac{J_1(\ell\,r)}{\ell\,r}\,P_\psi(\ell).
\end{eqnarray}
For a well-behaved $P_\psi(\ell)$, we can define, at zero separation $r = 0$, $C(0) := C_\perp(0) = C_\parallel(0)$ (i.e. the one-point variance of $\bfalp(\bfx)$ is finite).

For Gaussian random deflections, the mean magnification factor is given by an integral over a single image plane,
\begin{align}
\label{eq:muWinteg}
    \VEV{\mu_{\rm W}(\bfy)} & = \frac{1}{(2\pi)\,(C(0) + \sigma^2_{\rm W})}\, \int\,\rmd^2\bfx\,\exp\left[-\frac12\,\frac{|\bfx - \bfy - \bfalp_{\rm B}(\bfx)|^2}{C(0) + \sigma^2_{\rm W}}\right].
\end{align}
The second moment is given by an integral over double image planes,
\begin{eqnarray}
\label{eq:muW2dblinteg}
    \VEV{\mu_{\rm W}(\bfy)^2} & = & \int\,\rmd^2\bfx_1\, \int\,\rmd^2\bfx_2\,\frac{\exp\left[ -\frac12\,\textbf{u}^T(\bfx_1,\,\bfx_2;\,\bfy)\,\left( \textbf{C}(\bfr_{12}) + \sigma^2_{\rm W}\,\textbf{I} \right)^{-1}\,\textbf{u}(\bfx_1,\,\bfx_2;\,\bfy) \right]}{(2\pi)^2\,\sqrt{{\rm det}[ \textbf{C}(\bfr_{12}) + \sigma^2_{\rm W}\,\textbf{I}]}}.
\end{eqnarray}
To condense the expression, we have constructed a four-component vector
\begin{align}
    \textbf{u}\left( \bfx_1,\,\bfx_2;\,\bfy \right) = \left[\begin{array}{c}
x_{1,1} - y_1 - \alpha_{{\rm B}, 1}(\bfx_1) \\
x_{1,2} - y_2 - \alpha_{{\rm B}, 2}(\bfx_1) \\
x_{2,1} - y_1 - \alpha_{{\rm B}, 1}(\bfx_2) \\
x_{2,2} - y_2 - \alpha_{{\rm B}, 2}(\bfx_2) \\
\end{array}\right].
\end{align}
We also introduce the four-by-four identity matrix $\textbf{I}$, and define a four-by-four covariance matrix
\begin{align}
\label{eq:Cmatrix}
    \textbf{C}(\bfr) = \left[\begin{array}{cccc}
C(0) & 0 & C_\parallel(r)\,c^2 +  C_\perp(r)\,s^2  & C_\parallel(r)\,c\,s -  C_\perp(r)\,c\,s  \\
0 & C(0) & C_\parallel(r)\,c\,s -  C_\perp(r)\,c\,s   & C_\parallel(r)\,s^2 +  C_\perp(r)\,c^2 \\
C_\parallel(r)\,c^2 +  C_\perp(r)\,s^2  & C_\parallel(r)\,c\,s -  C_\perp(r)\,c\,s & C(0) & 0 \\
C_\parallel(r)\,c\,s -  C_\perp(r)\,c\,s   & C_\parallel(r)\,s^2 +  C_\perp(r)\,c^2 & 0 & C(0) \\
\end{array}\right],
\end{align}
where the shorthand notations $c:=\cos\varphi$ and $s:=\sin\varphi$ are used, under the polar coordinate parametrization $\bfr = r\,\left[ \cos\varphi,\,\sin\varphi \right]$. Given the matrix determinant ${\rm det}[\textbf{C}(\bfr)]=(C(0)+C_\parallel(r))\,(C(0)-C_\parallel(r))\,(C(0)+C_\perp(r))\,(C(0)-C_\perp(r))$, $\textbf{C}(\bfr)$ is a positive-definite matrix if and only if $C(0) \pm C_\parallel(r) >0$ and $C(0) \pm C_\perp(r) >0$. In our toy model, these are strictly guaranteed by \refeq{Caapara} and \refeq{Caaperp} provided that $P_\psi(\ell) > 0$. If the off-diagonal elements in \refeq{Cmatrix} were all vanishing, we would have concluded that $\VEV{\mu_{\rm W}(\bfy)^2} = \VEV{\mu_{\rm W}(\bfy)}^2$. Hence, the off-diagonal matrix elements of $\textbf{C}(\bfr)$ is the reason for the nonzero variance for $\mu_{\rm W}(\bfy)$. 

For giving numerical examples, we specify a simple analytic form for the potential power spectrum:
\begin{align}
    P_\psi(\ell) = \alpha^2_0\,\frac{\sigma^4_\psi}{(2\pi)^3}\,\exp\left( -\frac12\,\frac{\sigma^2_\psi\,\ell^2}{(2\pi)^2} \right).
\end{align}
The resultant root-mean-square (RMS) deflection is $\sqrt{\VEV{|\bfalp(\bfx)|^2}} = \sqrt{2}\,\alpha_0$. The parameter $\sigma_\psi$ is introduced to set the coherent scale of the deflections on the image plane. The deflection correlation functions can be explicitly computed, $C_\perp(r)=\alpha^2_0\,e^{-(1/2)\,(2\pi\,r/\sigma_\psi)^2}$, and $C_\parallel(r)=C_\perp(r)\,[1-(2\pi\,r/\sigma_\psi)^2]$. The RMS convergence is $\VEV{\kappa^2(\bfx)} = \VEV{[(1/2)\,\nabla^2\,\phi(\bfx)]^2} = \sqrt{2}\,(2\pi)\,\alpha_0/\sigma_\psi$.

To study the situation of a rapid varying $\mu_{\rm B}(\bfy)$, let us consider the background deflection of a fold caustic, which is the most commonly encountered caustic. We use the following parameterization for the fold caustic~\citep{1992grle.book.....S}, 
\begin{eqnarray}
    x_1 - \alpha_{\rm B, 1}(\bfx) & = & \frac12\,d_1\,x^2_1 + d_2\,x_1\,x_2 - \frac12\,d_1\,x^2_2, \\
    x_2 - \alpha_{\rm B, 2}(\bfx) & = & 2\,(1 - \kappa_0)\,x_2 + \frac12\,d_2\,x^2_1 - d_1\,x_1\,x_2 - \frac12\,d_2\,x^2_2.
\end{eqnarray}
The parameters are the local background convergence $\kappa_0$, and a gradient vector $\bfd=[d_1,\,d_2]$ which is related to the third-order derivative of the background lensing potential. We introduce $d=\sqrt{d^2_1 + d^2_2}$ and $\alpha=-\tan^{-1}(d_1/d_2)$ following the notation of \cite{2017ApJ...850...49V}. The coordinate system is conveniently chosen so that the caustic aligns with the $y_2$ axis on the source plane, and the corresponding critical curve on the image plane intersects the $x_1$ axis at an angle $\alpha$. As an example, we consider the case $d_1>0$ and $d_2=0$, so that $d=d_1$ and $\alpha=\pi/2$.

\begin{figure}[h]
    \centering
    \includegraphics[scale=0.43]{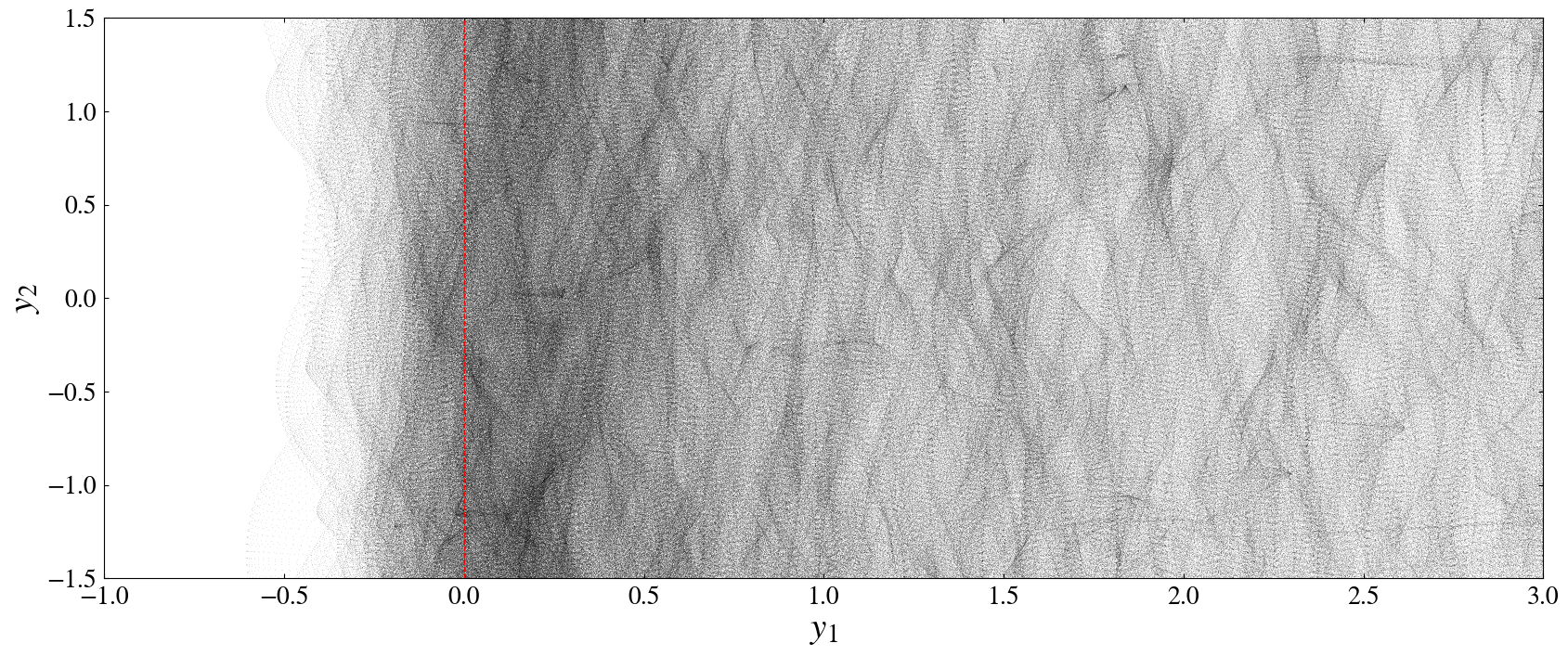}
    \caption{A corrugated micro caustic network that forms on the source plane in the vicinity of a background caustic (red dashed line). We adopt the model of Gaussian random deflections as introduced \refsec{gaussian}, setting parameter values $d=10^{-5}$, $\alpha_0 = 0.2$, and $\sigma_\psi=10$. The density of points is proportional to the magnification for a point source. Micro caustics are visible in the form of sharp linear features of high point densities.}
    \label{fig:micro_caustics}
\end{figure}

Several important scales can be identified on the source plane. First, the characteristic source size is $\sigma_{\rm W}$. Second, in the vicinity of the ideally smooth background caustic, the random deflections cannot be neglected. This is relevant within a source-plane width $\sim \alpha_0$ from the background caustic, as set by the root mean square (RMS) deflection. Within this proximity of the background caustic, the background magnification reaches $\mu_f \sim (\alpha_0\,d)^{-1/2}$, which we assume to be large. When $\mu_f$ is greater than the inverse of the typical convergence fluctuation $\sim (\alpha_0/\sigma_\psi)$, micro caustics join to form a network within this narrow band. The random deflections have a coherent scale $\sim \sigma_\psi$ on the image plane. When ray-traced onto the source plane following the background ray equation, this on average maps to a compressed scale $\sim \sigma_\psi/\mu_f = \sigma_\psi\,(\alpha_0\,d)^{1/2}$ for the corrugated micro caustic pattern, along the direction perpendicular to the background caustic. When the micro caustic network does arise, parametrically we have the hierarchy $\alpha_0 > \sigma_\psi\,(\alpha_0\,d)^{1/2}$. \reffig{micro_caustics} shows a numerical example of how a smooth macro caustic is replaced by a corrugated micro caustic pattern due to the Gaussian random deflections.

\begin{figure}[t]
    \centering
    \includegraphics[scale=0.5]{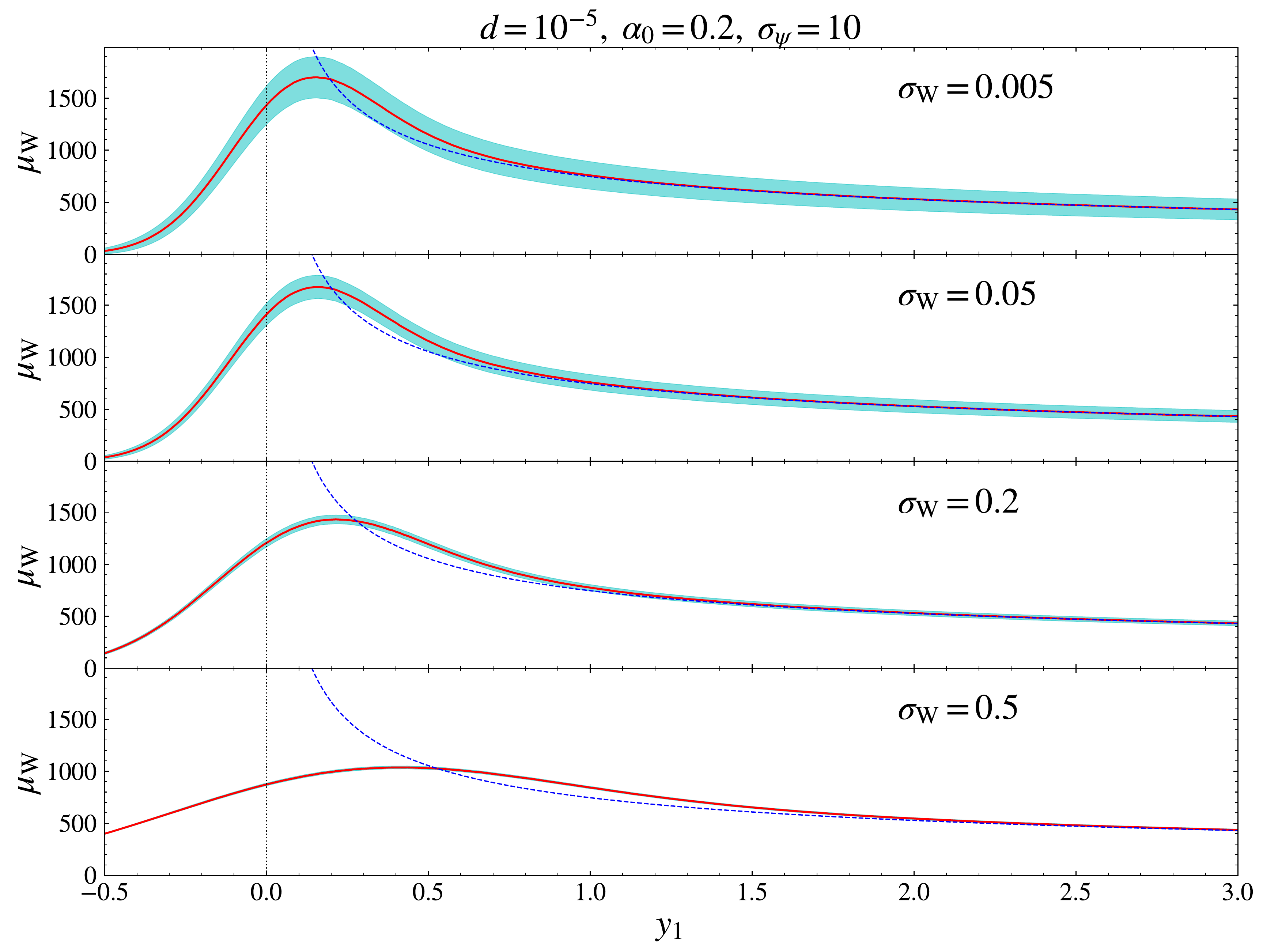}
    \caption{Total magnification factor $\mu_{\rm W}$ for a finite source with a Gaussian profile as a function of distance to the background caustic (vertical dotted line at $y_1 = 0$). We set parameter values $\kappa_0=0.7$, $d=10^{-5}$, $\alpha_0=0.2$ and $\sigma_\psi=10$, as defined in the text, and consider increasing source sizes $\sigma_{\rm W}=0.005,\,0.05,\,0.2,\,0.5$ (from top to bottom). We show both the mean magnification $\VEV{\mu_{\rm W}}$ (red solid) and its standard deviation $\sqrt{\VEV{\mu^2_{\rm W}}-\VEV{\mu_{\rm W}}^2}$ due to random deflections (cyan band; $\pm 1\,\sigma$). The magnification factor for an idealized point source without random deflections is also shown for comparison (blue dashed). For $\sigma_{\rm W} < \alpha_0$, the height and width of the peak $\VEV{\mu_{\rm W}}$ are set by $\alpha_0$, while the magnification fluctuations decreases as $\sigma_{\rm W}$ increases. For $\sigma_{\rm W} > \alpha_0$, the height and width of the peak $\VEV{\mu_{\rm W}}$ are instead set by $\sigma_{\rm W}$, and the magnification fluctuations are highly smoothed out.}
    \label{fig:mu_example}
\end{figure}

We evaluate the multi-dimensional integrations \refeq{muWinteg} and \refeq{muW2dblinteg} using the widely used Monte Carlo algorithm \texttt{vegas}~\citep{Lepage1978vegas, vegasEnhanced}. In \reffig{mu_example}, we present numerical examples that verify the intuitions we have developed based on our analytic scrutiny of the Gaussian random deflection model. We see that the mean magnification $\VEV{\mu_{\rm W}(\bfy)}$ as a function of source center $\bfy$ indeed has a smoothed behavior at the background caustic. The maximal mean magnification and the width of the smoothed curve is set by the size of random deflections $\alpha_0$ if that exceeds the source size $\alpha_0 > \sigma_{\rm W}$. From one realization to another, $\mu_{\rm W}(\bfy)$ fluctuates around the mean $\VEV{\mu_{\rm W}(\bfy)}$. However, such fluctuations are suppressed when the source size $\sigma_{\rm W}$ is larger than the characteristic separation of micro caustics.

\section{Random deflections from compact microlenses}
\label{sec:microlens}

\begin{figure}[h]
    \centering
    \includegraphics[scale=0.43]{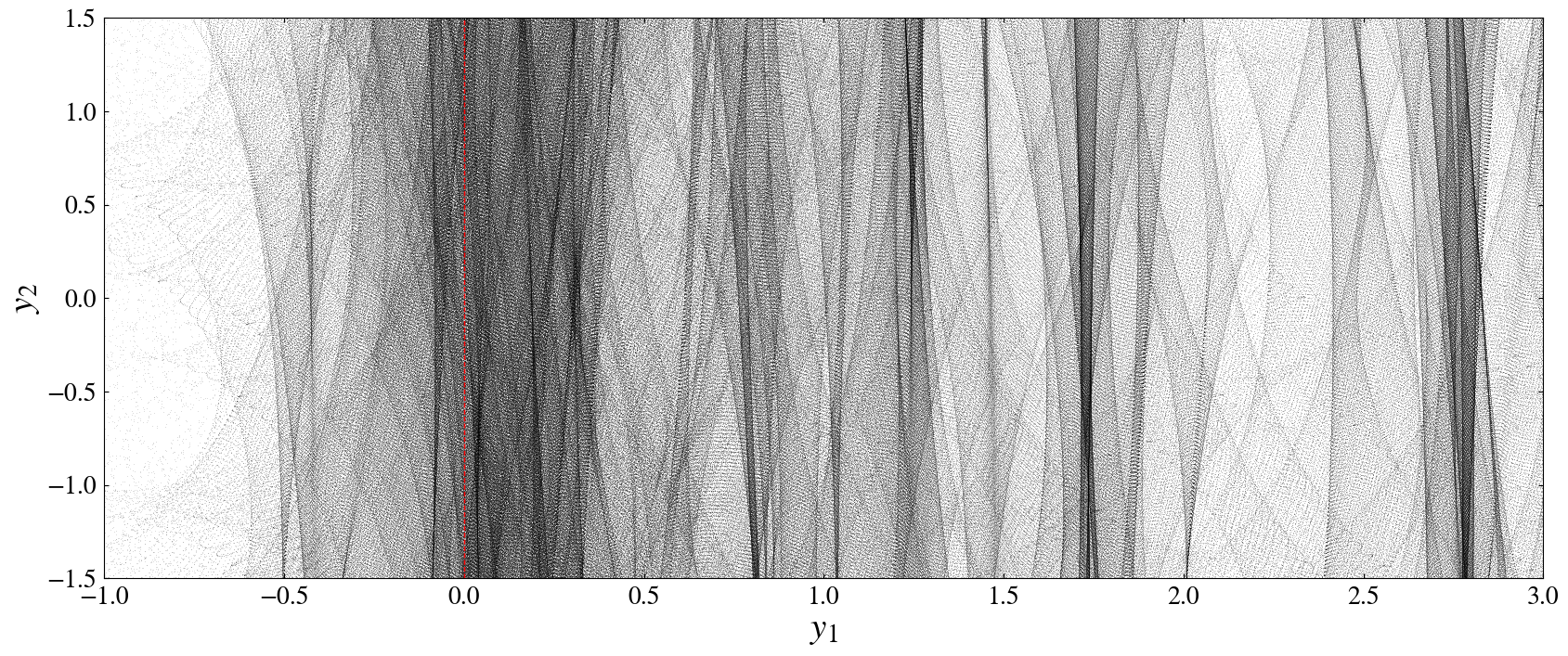}
    \caption{The corrugated micro caustic network caused by point microlenses with $\theta_\star=1$ and $\kappa_\star=0.02$ ($\sigma_{\rm eff} = 0.14$; $\sigma_{\rm ml}$ comparable to $\alpha_0$ used for \reffig{micro_caustics}). Parameter choices for the background fold caustic are the same as adopted for \reffig{micro_caustics}. In comparison to the case of Gaussian random deflections, the micro caustics induced by point lenses are stronger, and lie preferentially parallel to the background caustic.}
    \label{fig:micro_caustics_ml}
\end{figure}

In most astrophysical situations, random deflections are due to compact lenses such as individual stars. \reffig{micro_caustics_ml} shows an example of the corrugated micro caustic network cast by random point lenses, which looks qualitatively different from that formed due to Gaussian random deflections.

An important difference between the case of point microlenses and that of Gaussian random deflections is that very large deflections are generated with a low but non-negligible probability. This happens when the ray encounters a microlens at a small impact parameter. This not only greatly enhances the flux fluctuations, but also renders the distribution of $\bfalp(\bfx)$ heavy-tailed~\citep{Katz1986RandomScattering}. As a consequence, $\bfalp(\bfx)$ has divergent second-order and higher-order moments, which hinders exact analytic treatment.

For a simple discussion, we assume that all microlenses have the same mass, and hence the same angular Einstein scale $\theta_\star$. We will comment on the generalization to the case of a continuous microlens mass distribution toward the end of the this Section. Other angular scales of the problem can all be expressed in units of $\theta_\star$. We assume that the microlenses are uniformly distributed on the lens plane and contribute a mean convergence $\kappa_\star > 0$.

The microlensing deflection can be written as
\begin{align}
    \bfalp_{\rm ml}(\bfx) = - \kappa_\star\,\bfx + \theta^2_\star\,\sum^N_{I=1}\,\frac{\bfx - \bfz_I}{|\bfx - \bfz_I|^2}.
\end{align}
In the second term, we sum over contributions from all microlenses $I=1,2,\cdots,\,N$, which are located at $\bfz_I$'s, respectively. In the first term, we subtract the mean deflection due to a uniform mass sheet of convergence $\kappa_\star$. Including the first term enforces that $\VEV{\bfalp_{\rm ml}(\bfx)}=0$. Unlike in the case of Gaussian random deflections, analytic calculations of the CFs for $\bfalp_{\rm ml}(\bfx)$ are difficult for the case of point microlenses. Still, the CFs can be reduced to those corresponding to a single microlens, provided that microlenses have independent positions on the lens plane. This forms the basis of the following analytic treatment.

\subsection{One-point deflection statistics}

\begin{figure}[t]
    \centering
    \includegraphics[scale=0.4]{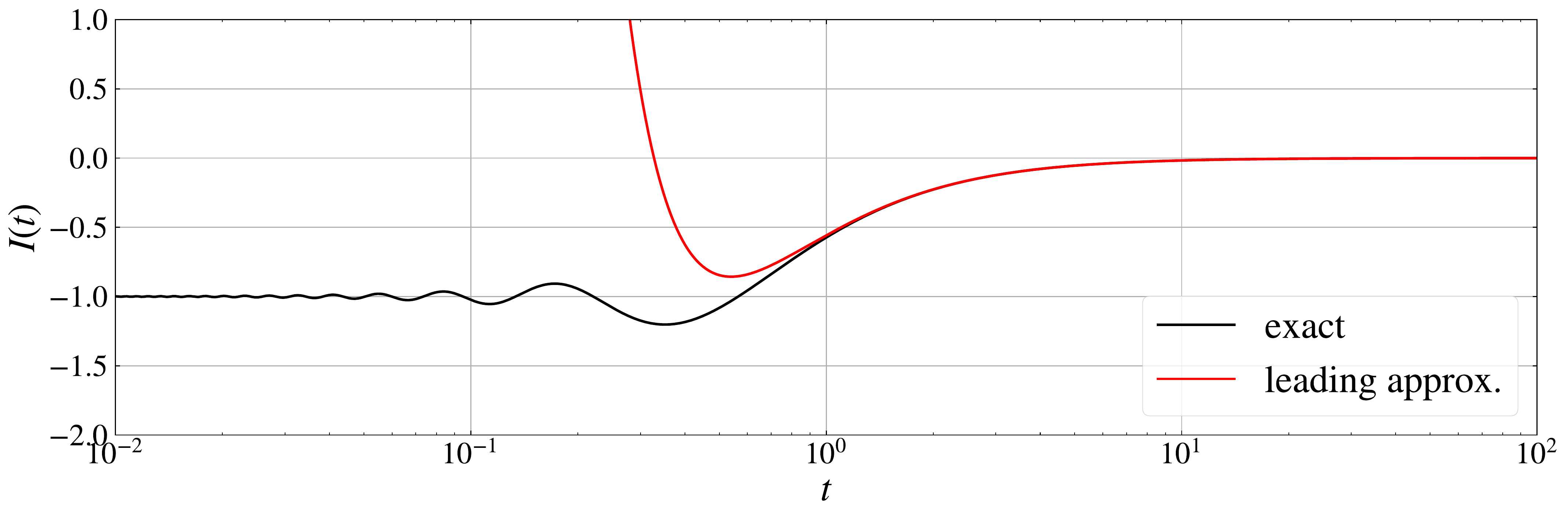}
    \caption{The function $I(t)$ as defined in \refeq{It}. We compare the exact numerical evaluation (black) and the leading-order approximation (red), $-(1/2\,t^2)\,(1 - \gamma_E + \ln\,2 t)$, in the $(1/t)$-expansion. The leading-order approximation works remarkably well for $t \gtrsim 1$.}
    \label{fig:Ifun}
\end{figure}

Let $P_1[\bfalp_{\rm ml}(\bfx)]$ be the PDF for $\bfalp_{\rm ml}(\bfx)$. Its Fourier transform gives the one-point CF:
\begin{align}
    \Phi_1[\bfl;\,\bfx] = \VEV{e^{i\,\bfl\cdot\bfalp_{\rm ml}(\bfx)}} = e^{-i\,\kappa_\star\,\bfl\cdot\bfx}\,e^{ N\,\ln\varphi_1[\bfl;\,\bfx] }.
\end{align}
Here $\varphi_1[\bfl;\,\bfx]$ is the one-point CF due to a single microlens. We specify that each microlens is uniformly distributed within a disk of radius $R$, and compute for $\bfalp_{\rm ml}(\bfzero)$. We perform the following integral (\reffig{ml_defl_integs}; left)
\begin{align}
\label{eq:Itinteg}
    \ln\varphi_1[\bfl] \equiv \ln\varphi_1[\bfl;\,\bfzero] & \approx \varphi_1[\bfl;\,\bfzero] - 1 = \frac{1}{\pi\,R^2}\,\int\displaylimits_{|\bfz| < R}\,\rmd^2\bfz\,\left( e^{ -i\,\theta^2_\star\,\bfl\cdot\bfz/|\bfz|^2 } - 1 \right) 
    \equiv I\left( \frac{R}{\theta^2_\star\,\ell} \right),
\end{align}
where $\ell = |\bfell|$. The integral $I(t)$ can be expressed as a series expansion:
\begin{align}
\label{eq:It}
    I(t) & \equiv \frac{2}{t^2}\,\int^{t}_0\,t'\,\rmd t'\,\left[ J_0\left(\frac{1}{t'} \right) - 1 \right] \nonumber\\
    & = - \frac{t^{-2}}{2}\,\left( 1 - \gamma_E + \ln\,2\,t \right) - \frac{t^{-4}}{64} + \frac{t^{-6}}{4608} + \frac{t^{-8}}{884736} +  \cdots,
\end{align}
where $J_0(x)$ is the Bessel function of the first kind, and $\gamma_E \approx 0.577216$ is the Euler–Mascheroni constant. For sufficiently large $R$, $t \gtrsim 1$, and we observe that the series rapidly converges. Keeping only the $\mathcal{O}(1/t^2)$ term gives a decent approximation (\reffig{Ifun}):
\begin{align}
\label{eq:lnvphi1}
    \ln\varphi_1[\bfl] \approx -\frac12\,\frac{\theta^4_\star\,\ell^2}{R^2}\,\left( 1 - \gamma_E + \ln\frac{2\,R}{\theta^2_\star\,\ell} \right).
\end{align}
The presence of the $\ell$-dependent logarithm $\ln(2\,R/\theta^2_\star\,\ell)$ reflects that the deflection due to one microlens follows a heavy-tailed distribution; a Taylor expansion in $\bfell$ around $\bfell = 0$ is not possible.

Using $\kappa_\star = N\,\theta^2_\star/R^2$, and setting $\bfx=\bfzero$, we derive
\begin{align}
\label{eq:Phi1}
    \Phi_1[\bfell] = \exp\left[ - \frac12\,|\bfell|^2\,\kappa_\star\,\theta^2_\star\,\left( 1 - \gamma_E + \ln\frac{2\,R}{\theta^2_\star\,\ell} \right) \right].
\end{align}
Note that in the large $N$ limit, the higher order terms in \refeq{It}, i.e. $\sim t^{-2\,k}$ for $k=2,3,\cdots$, are suppressed by $\sim 1/N^{k-1}$, respectively.

The variance of the deflection angle $\VEV{|\bfalp_{\rm ml}(\bfzero)|^2}\approx \kappa_\star\,\theta^2_\star\,(1-\gamma_E + \ln(2\,R/\theta^2_\star\,\ell))$ is always of order $\sigma^2_{\rm eff} = \theta^2_\star\,\kappa_\star$. Additionally, there is a multiplicative factor that depends logarithmically on the IR cutoff scale $R$ as well as the correlation scale of interest $\sim 1/\ell$. By fixing $R=R_*$ and $\ell=\ell_*$ in the logarithm for some appropriate values of $R_*$ and $\ell_*$, we essentially approximate $P_1[\bfalp_{\rm ml}]$ as a two-dimensional Gaussian distribution:
\begin{align}
\label{eq:P1gaussian}
    P_1[\bfalp_{\rm ml}] \approx \frac{1}{2\pi\,\sigma^2_{\rm ml}(R_*,\,\ell_*)}\,\exp\left( - \frac12\,\frac{|\bfalp_{\rm ml}|^2}{\sigma^2_{\rm ml}(R_*,\,\ell_*)} \right),
\end{align}
where we introduce
\begin{align}
\label{eq:sigma2ml}
    \sigma^2_{\rm ml}(R_*,\,\ell_*) = \theta^2_\star\,\kappa_\star\,\left( 1 - \gamma_E + \ln\frac{2\,R_*}{\theta^2_\star\,\ell_*} \right).
\end{align}
The key question is what the appropriate values for $R_*$ and $\ell_*$ would be under this approximation.

The $R$ dependence originates from the collective influence of faraway microlenses. One straightforward choice for $R_*$ would be to use the full angular scale on the image plane over which the microlens convergence has nearly a uniform value $\kappa_\star$. This scale can be orders of magnitude larger than the Einstein scale $\theta_\star$. For galaxy or intracluster microlensing, it may be taken to be the characteristic extent of the galactic stellar halo or the intracluster light halo, respectively. 

\cite{Katz1986RandomScattering} (hereafter KBP86) instead propose that $R_*$ should be the scale over which the microlenses produce incoherent deflections across the distribution of micro images, arguing that coherent deflections, for any fixed microlens realization, lead to an overall uniform shift of the micro images without affecting the magnification. Define an effective source size
\begin{align}
\label{eq:sigmaeff}
    \sigma_{\rm eff} = \sqrt{\sigma^2_{\rm W} + \kappa_\star\,\theta^2_\star},
\end{align}
which accounts for intrinsic source size and effective ``broadening'' due to random microlensing deflections. Let $\mu_{\rm B}$ be the macro magnification near a background fold caustic as introduced in \refsec{gaussian}. The maximal value of $\mu_{\rm B}$ is limited by the effective source size $\sigma_{\rm eff}$, reaching $\mu_f \sim (\sigma_{\rm eff}\,d)^{-1/2}$. Either macro image thus has an extent $\sim \mu_{\rm B}\,\sigma_{\rm eff}$. Following the choice of KBP86, we may set $R_*$ to $\mu_{\rm B}\,\sigma_{\rm eff} \lesssim \mu_f\,\sigma_{\rm eff}= (\sigma_{\rm eff}/d)^{1/2}$.

However, the choice of KBP86 is not entirely justifiable if the background lens model is non-uniform. The issue can be particularly non-trivial near a macro caustic where the (small) gradients of background convergence and shear set the strength of the caustic --- large-scale Poisson fluctuations in the microlens number not only generate a coherent deflection, but also contribute small gradients of convergence and shear which can modify the effective caustic strength from realization to realization. In this case, we set $R_*$ to be the largest scale over which the microlenses are uniformly distributed.

The typical value of $\ell_*$ to be set in the logarithm can be related to the inverse of the width of $P^{\rm W}_1[\bfalp]$, which is the amount needed to compensate for $\bfx - \bfy - \bfalp_{\rm B}(\bfx) \neq 0$. This means that we can set $\ell_*=1/\sigma_{\rm eff}$.

For the choice $R_*=\mu_f\,\sigma_{\rm eff}$ (KBP86) and $\ell_*=1/\sigma_{\rm eff}$, the truncation of expansion \refeq{It} is valid if
\begin{align}
\label{eq:th2ltoR}
    \frac{\theta^2_\star\,\ell_*}{R_*} = \frac{\theta^2_\star}{\mu_{\rm B}\,\sigma^2_{\rm eff}} = \frac{\theta_\star\,\kappa^{-1/2}_\star\,\mu^{-1}_{\rm B}}{\sigma_{\rm eff}}\,\frac{\theta_\star\,\kappa^{1/2}_\star}{\sigma_{\rm eff}} \lesssim 1.
\end{align}
If $\sigma_{\rm eff}$ is larger than the characteristic scale of the micro-caustic network $\sim \theta_\star\,\kappa^{-1/2}_\star\,\mu^{-1}_{\rm B}$, then the first factor is smaller than unit. This corresponds to the situation where either the physical source extent $\sigma_{\rm W}$ overlaps multiple micro caustics, or the typical microlensing broadening $\theta_\star\,\kappa^{1/2}_\star$ overlaps multiple micro caustics (the regime of many micro images). The second factor is no larger than unity. Hence, $\theta^2_\star\,\ell_*/R_* < 1$ is satisfied in this situation.

\begin{figure}[t]
    \centering
    \includegraphics[scale=0.4]{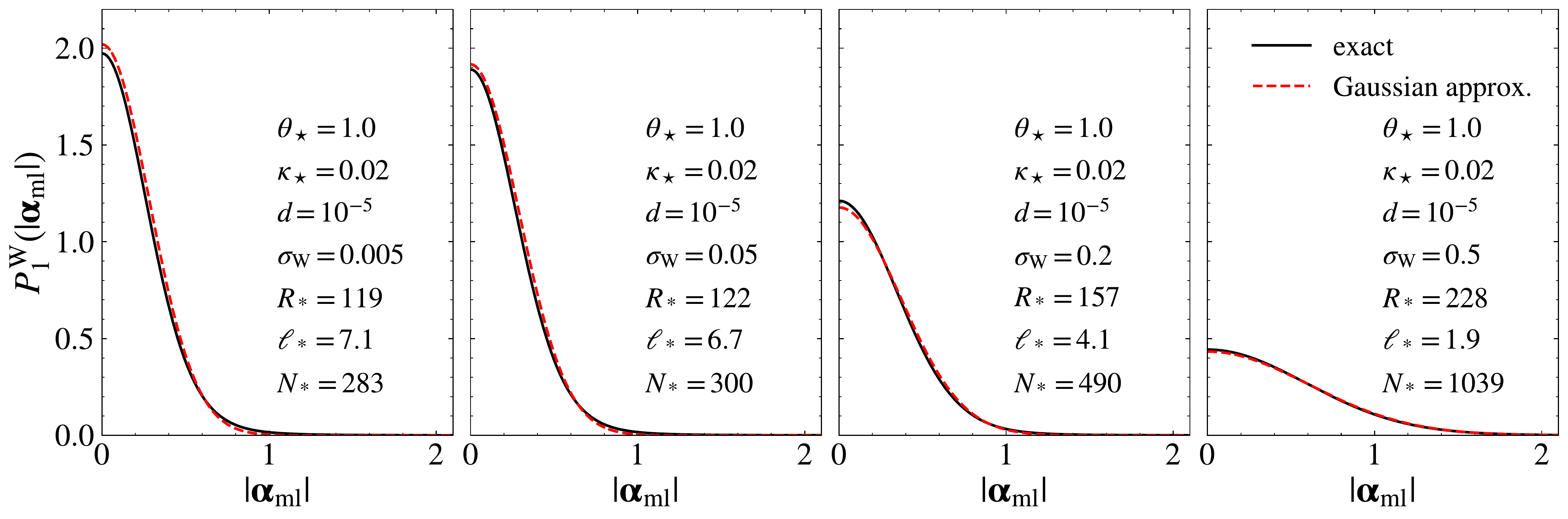}
    \caption{Source-profile convoluted distribution $P^{\rm W}_1(\bfalp_{\rm ml})=P^{\rm W}_1(|\bfalp_{\rm ml}|)$. We consider the case in which the highest possible mean magnification $\VEV{\mu_{\rm W}(\bfy)}$ is realized when the source center $\bfy$ is within a distance $\sim \sigma_{\rm eff}$ (see \refeq{sigmaeff}) of the background caustic, and set $R_*=(\sigma_{\rm eff}/d)^{1/2}$ due to \cite{Katz1986RandomScattering}. Panels correspond to different values for the characteristic source size $\sigma_{\rm W}$. Good agreement is found between direct numerical evaluation of \refeq{P1Wdef} (black solid curves), and our analytic model (\refeq{P1gaussian}; red dashed curves) which approximates the PDF for the microlensing random deflections, $P _1(|\bfalp_{\rm ml}|)$, as an isotropic two-dimensional Gaussian distribution.}
    \label{fig:P1W}
\end{figure}

As we verify numerically in \reffig{P1W}, assuming the random microlensing deflections to follow approximately a two-dimensional Gaussian distribution can fairly accurately reproduce the exact source-profile convoluted distribution $P^{\rm W}_1(\bfalp_{\rm ml})$, provided that $\theta^2_\star\,\ell_* < R_*$.

\refeq{sigma2ml} can be recast into the following form
\begin{align}
    \sigma^2_{\rm ml}(R_*,\,\ell_*) = \theta^2_\star\,\kappa_\star\,\left[ \ln\left( 2\,e^{1-\gamma_E} \, N^{1/2}_* \right) - \ln\left(\theta_\star\,\kappa^{1/2}_\star\,\ell_* \right) \right],
\end{align}
where $N_* = \kappa_\star\,R^2_*/\theta^2_\star$ is the number of microlenses within an image-plane disk of radius $R_*$. The first logarithmic term has been previously derived by KBP86 as $\ln(3.05\,N^{1/2}_*)$. Our result includes a second term $\ln(\theta_\star\,\kappa^{1/2}_\star\,\ell_*)$ not found in KBP86, as we have argued that it is appropriate to set $\ell_* = 1/\sigma_{\rm eff}$ (see \refeq{sigmaeff}). For small source sizes, $\sigma_{\rm W} < \theta_\star\,\kappa^{1/2}_\star$ and hence $1/\ell_* = \sigma_{\rm eff} \approx \theta_\star\,\kappa^{1/2}_\star$, rendering this second logarithm numerically negligible. On the other hand, this second logarithm increases the effective Gaussian width for large sources, i.e. $\sigma_{\rm eff} \approx \sigma_{\rm W} > \theta_\star\,\kappa^{1/2}_\star$.

From the above analysis, we obtain the following approximation formula for the mean magnification factor,
\begin{align}
\label{eq:muWvevML}
    \VEV{\mu_{\rm W}(\bfy)} & \approx \frac{1}{(2\pi)\,(\sigma^2_{\rm ml}(R_*,\,\ell_*) + \sigma^2_{\rm W})}\, \int\,\rmd^2\bfx\,\exp\left[-\frac12\,\frac{|\bfx - \bfy - \bfalp_{\rm B}(\bfx)|^2}{\sigma^2_{\rm ml}(R_*,\,\ell_*) + \sigma^2_{\rm W}}\right],
\end{align}
where the value of $\sigma^2_{\rm ml}(R_*,\,\ell_*)$ must be appropriately set using \refeq{sigma2ml}. 

\subsection{Two-point deflection statistics}

Apart from the mean magnification $\VEV{\mu_{\rm W}(\bfy)}$, we would like to know the statistical fluctuations in the magnification factor between random microlens realizations. As we have shown in \refeq{muWy2}, calculating the second moment $\VEV{\mu_{\rm W}(\bfy)^2}$ requires the knowledge of the two-point CF for microlensing deflection:
\begin{align}
    \Phi_2[\bfl_1,\,\bfl_2;\,\bfx_1,\,\bfx_2] = \VEV{e^{i\bfl_1\cdot\bfalp_{\rm ml}(\bfx_1)}\,e^{i\bfl_2\cdot\bfalp_{\rm ml}(\bfx_2)}} = e^{-i\,\kappa_\star\,(\bfl_1\cdot\bfx_1 + \bfl_2\cdot\bfx_2 )}\,\exp\left[N\,\ln\varphi_2\left[ \bfl_1,\,\bfl_2;\,\bfx_1,\,\bfx_2 \right]\right].
\end{align}
Here $\varphi_2\left[ \bfl_1,\,\bfl_2;\,\bfx_1,\,\bfx_2 \right]$ is the two-point CF due to a single microlens, and is given by the following integral
\begin{eqnarray}
\label{eq:lnvphi2}
    \ln\varphi_2\left[ \bfl_1,\,\bfl_2;\,\bfx_1,\,\bfx_2 \right] & \approx & \varphi_2\left[ \bfl_1,\,\bfl_2;\,\bfx_1,\,\bfx_2 \right] - 1 \nonumber\\
    & = & \frac{1}{\pi\,R^2}\,\int\displaylimits_{|\bfz|<R}\,\rmd^2\bfz\,\left[ \exp\left( i\,\theta^2_\star\,\left( \frac{\bfl_1\cdot(\bfx_1 - \bfz)}{|\bfx_1 - \bfz|^2} + \frac{\bfl_2\cdot(\bfx_2 - \bfz)}{|\bfx_2 - \bfz|^2} \right) \right) - 1 \right].
\end{eqnarray}

A closed-form result for \refeq{lnvphi2} for arbitrary wave vectors $\bfl_1$ and $\bfl_2$ and arbitrary image positions $\bfx_1$ and $\bfx_2$ is unknown to us. For finite sources, nevertheless, it is useful to analytically extract the part that has quadratic dependence on $\bfl_1$ and $\bfl_2$ (with additional logarithms as a result of non-analyticity). Keeping only this quadratic part amounts to approximating the two-point joint PDF for the random deflections
\begin{align}
    P_2\left(\bfalp_{\rm ml}(\bfx_1),\,\bfalp_{\rm ml}(\bfx_2) \right) = \int\,\frac{\rmd^2\bfl_1}{(2\pi)^2}\,\int\,\frac{\rmd^2\bfl_2}{(2\pi)^2}\,e^{-i\,\bfl_1\cdot\bfalp_{\rm ml}(\bfx_1)}\,e^{-i\,\bfl_2\cdot\bfalp_{\rm ml}(\bfx_2)}\,\Phi_2\left[ \bfl_1,\,\bfl_2;\,\bfx_1,\,\bfx_2 \right],
\end{align}
as a two-dimensional normal distribution, which will lead to analytic simplification of the expression for $\VEV{\mu_{\rm W}(\bfy)^2}$. We argue that this approximation is reasonable if the effective source size overlaps multiple micro-caustics $\theta_\star/(\kappa^{1/2}_\star\,\mu_{\rm B}\,\sigma_{\rm eff}) \lesssim 1$, which, as we have argued before with \refeq{th2ltoR}, necessarily implies that $\theta^2_\star\,\ell_*/R_* \lesssim 1$. In this regime, $\bfl_1$ and $\bfl_2$ are typically on the order of $\ell_*$, and the separation between the two image-plane positions is typically comparable to the macro image extent $r_{12} \equiv |\bfx_1 - \bfx_2|$. With the condition $\theta^2_\star\,\ell_{1,2}/|\bfx_1 - \bfx_2| < 1$, an analytic approximation for the integral \refeq{lnvphi2} can be obtained. This is given by \refeq{lnvphi2approx}. See \refapp{twoptinteg} for the derivation. The result is the following approximation for the two-point CF:
\begin{eqnarray}
\label{eq:lnPhi2}
    \ln\Phi_2[\bfl_1,\,\bfl_2;\,\bfx_1,\,\bfx_2] & \approx & - \frac{\kappa_\star\,\theta^2_\star}{2}\,\left( |\bfl_1|^2 + |\bfl_2|^2 \right)\,\left( 1 - \gamma_E + \ln\frac{2\,R_*}{\theta^2_\star\,\ell_*} \right) \\
    && - \kappa_\star\,\theta^2_\star\,\left[ \left( \ln\frac{2\,R_*}{r_{12}} + \frac12\,\ln\left( 1 + \frac{r^2_{12}}{4\,R^2_*} \right) - \ln 2 + \frac12 \right)\,\left( \bfl_1\cdot\bfl_2 \right) - \frac{\left(\bfl_1\cdot\bfr_{12}\right)\,\left(\bfl_2\cdot\bfr_{12}\right)}{r^2_{12}}\right]. \nonumber
\end{eqnarray}
Terms of $|\bfl_1|^2$ and $|\bfl_2|^2$ encode auto correlation at a single image-plane point and are consistent with the one-point CF in \refeq{Phi1}, while the cross terms encode cross correlation between a pair of image-plane points. Since the cross terms only depend on $\bfr_{12}=\bfx_2 - \bfx_1$, but not on $\bfx_1$ and $\bfx_2$ separately, \refeq{lnPhi2} preserves the expected symmetries of the two-point microlensing deflection statistics under spatial translations and rotations on the image plane.

Through a direct calculation of $\langle \alpha_{{\rm ml}, i}(\bfx_1)\,\alpha_{{\rm ml}, j}(\bfx_2) \rangle$, we verify that the second line of \refeq{lnPhi2} can be rewritten as
\begin{align}
     - \left[ C^{\rm ml}_{\parallel}(r_{12};\,R_*)\,\frac{\left(\bfl_1\cdot\bfr_{12}\right)\,\left(\bfl_2\cdot\bfr_{12}\right)}{r^2_{12}} + C^{\rm ml}_{\perp}(r_{12};\,R_*)\,\left( \bfl_1\cdot\bfl_2 - \frac{\left(\bfl_1\cdot\bfr_{12}\right)\,\left(\bfl_2\cdot\bfr_{12}\right)}{r^2_{12}} \right) \right].
\end{align}
where $C^{\rm ml}_{\parallel}(r;\,R_*)$ and $C^{\rm ml}_{\perp}(r;\,R_*)$ are respectively the two-point correlation functions for random microlensing deflections parallel and perpendicular to the separation vector (as defined via decomposition \refeq{2ptalpalp}):
\begin{eqnarray}
\label{eq:CL}
C^{\rm ml}_\parallel(r;\,R_*) & = & \kappa_\star\,\theta^2_\star\,\left[ \ln\frac{2\,R_*}{r} + \frac12\,\ln\left( 1 + \frac{r^2}{4\,R^2_*} \right) - \ln 2 - \frac12 \right], \\
\label{eq:CT}
C^{\rm ml}_\perp(r;\,R_*) & = & \kappa_\star\,\theta^2_\star\,\left[ \ln\frac{2\,R_*}{r} + \frac12\,\ln\left( 1 + \frac{r^2}{4\,R^2_*} \right) - \ln 2 + \frac12 \right].
\end{eqnarray}
Both functions are sensitive to the ``infrared'' cutoff scale $R_*$ and are logarithmically divergent in the limit $r \rightarrow 0$, so the correlation at zero separation is ill-defined. This is unlike our toy model of Gaussian random deflections with a regular lensing potential power spectrum, as we have studied in \refsec{gaussian}, for which $C_\parallel(0) = C_\perp(0)$ is finite. To our knowledge, \refeq{CL} and \refeq{CT} have not been presented before in the literature.

Based on \refeq{lnPhi2}, we derive an approximate formula for $\VEV{\mu_{\rm W}(\bfy)^2}$ in a form identical to \refeq{muW2dblinteg}:
\begin{eqnarray}
\label{eq:muW2dblintegML}
    \VEV{\mu_{\rm W}(\bfy)^2} & = & \int\,\rmd^2\bfx_1\, \int\,\rmd^2\bfx_2\,\frac{\exp\left[ -\frac12\,\textbf{u}^T(\bfx_1,\,\bfx_2;\,\bfy)\,\left( \textbf{C}_{\rm ml}(\bfr_{12}) + \sigma^2_{\rm W}\,\textbf{I} \right)^{-1}\,\textbf{u}(\bfx_1,\,\bfx_2;\,\bfy) \right]}{(2\pi)^2\,\sqrt{{\rm det}[ \textbf{C}_{\rm ml}(\bfr_{12}) + \sigma^2_{\rm W}\,\textbf{I}]}}.
\end{eqnarray}
where we introduce the covariance matrix appropriate for random point microlenses:
\begin{align}
\label{eq:covCml}
    \textbf{C}_{\rm ml}(\bfr) = \left[\begin{array}{cccc}
C^{\rm ml}(0) & 0 & C^{\rm ml}_\parallel(r)\,c^2 +  C^{\rm ml}_\perp(r)\,s^2  & C^{\rm ml}_\parallel(r)\,c\,s -  C^{\rm ml}_\perp(r)\,c\,s  \\
0 & C^{\rm ml}(0) & C^{\rm ml}_\parallel(r)\,c\,s -  C^{\rm ml}_\perp(r)\,c\,s   & C^{\rm ml}_\parallel(r)\,s^2 +  C^{\rm ml}_\perp(r)\,c^2 \\
C^{\rm ml}_\parallel(r)\,c^2 +  C^{\rm ml}_\perp(r)\,s^2  & C^{\rm ml}_\parallel(r)\,c\,s -  C^{\rm ml}_\perp(r)\,c\,s & C^{\rm ml}(0) & 0 \\
C^{\rm ml}_\parallel(r)\,c\,s -  C^{\rm ml}_\perp(r)\,c\,s   & C^{\rm ml}_\parallel(r)\,s^2 +  C^{\rm ml}_\perp(r)\,c^2 & 0 & C^{\rm ml}(0) \\
\end{array}\right].
\end{align}
Here on the diagonal we use
\begin{align}
    C^{\rm ml}(0) := \kappa_\star\,\theta^2_\star\,\left( 1 - \gamma_E + \ln\frac{2\,R_*}{\theta^2_\star\,\ell_*} \right),
\end{align}
which depends on the choice for $R_*$ and $\ell_*$. Note that $C^{\rm ml}(0)$ is not given by the $r \rightarrow 0$ limit of $C^{\rm ml}_\parallel(r)$ or $C^{\rm ml}_\perp(r)$. 

The covariance matrix $\textbf{C}(\bfr)$ being positive definite requires $C^{\rm ml}(0) \pm C^{\rm ml}_\parallel(r) > 0$ and $C^{\rm ml}(0) \pm C^{\rm ml}_\perp(r) > 0$. Unlike in the Gaussian random deflection model, this is not strictly guaranteed here, which is a shortcoming of our analytic approximation. However, violation occurs in two regimes, $r \gg R_*$ or $r \ll \theta^2_\star\,\ell_*$, both of which are not expected to have important contributions to $\VEV{\mu_{\rm W}(\bfy)^2}$. In practice, we must regularize the logarithmic divergences in order for the integral \refeq{muWy2} to be well behaved. For an example, we present in \refapp{Creg} one regularization scheme, which, as we numerically test out, renders the result for $\VEV{\mu_{\rm W}(\bfy)^2}$ insensitive to regularization parameter choices.

Our results are readily applicable to the special case of uniform background convergence and shear. This situation has already been intensively studied, mostly in the context of quasar microlensing. In \refapp{uniformbkg}, we derive additional analytic results for this special case, and remark on comparisons to the literature.

While \refeq{muW2dblintegML} assumes a Gaussian source, \refeq{lnPhi2}, \refeq{CL} and \refeq{CT} are generally valid independent of the specific source profile. For stellar photospheres, the uniform disk would be a more appropriate source model than the Gaussian one. However, the $\bfell_1$- and $\bfell_2$-integrals cannot be analytically carried out, unlike for a Gaussian source, which is a shortcoming of our results. 

\subsection{Numerical experiments}

We validate the semi-analytic results we have derived for $\VEV{\mu_{\rm W}(\bfy)}$ and $\VEV{\mu_{\rm W}(\bfy)^2}$ using numerical experiments. We set parameters $\theta_\star=1$ and $d=10^{-5}$ as adopted in \reffig{micro_caustics_ml}. Random microlenses with identical $\theta_\star$ are generated within a circular disk that centers on the macro critical curve and has a radius $R_*=1500$. We efficiently compute the summed deflection from a large number of microlenses using the hierarchical tree algorithm~\citep{Wambsganss1999MLTreeAlgorithm}. For $\kappa_\star=0.004,\,0.02,\,0.1$ that we simulate, we need to include for each realization $N \approx 9000,\,45000,\,225000$ point lenses, respectively. We sample the image-plane vicinity of the macro critical curve with a large number of rays. These rays are inversely traced onto the source plane~\citep{KayserRefsdalStabell1986microlensing, Wambsganss1992muPDFML}, which can then be used to calculate $\mu_{\rm W}(\bfy)$ for any source profile and central position.

\begin{figure}[t]
    \centering
    \includegraphics[scale=0.53]{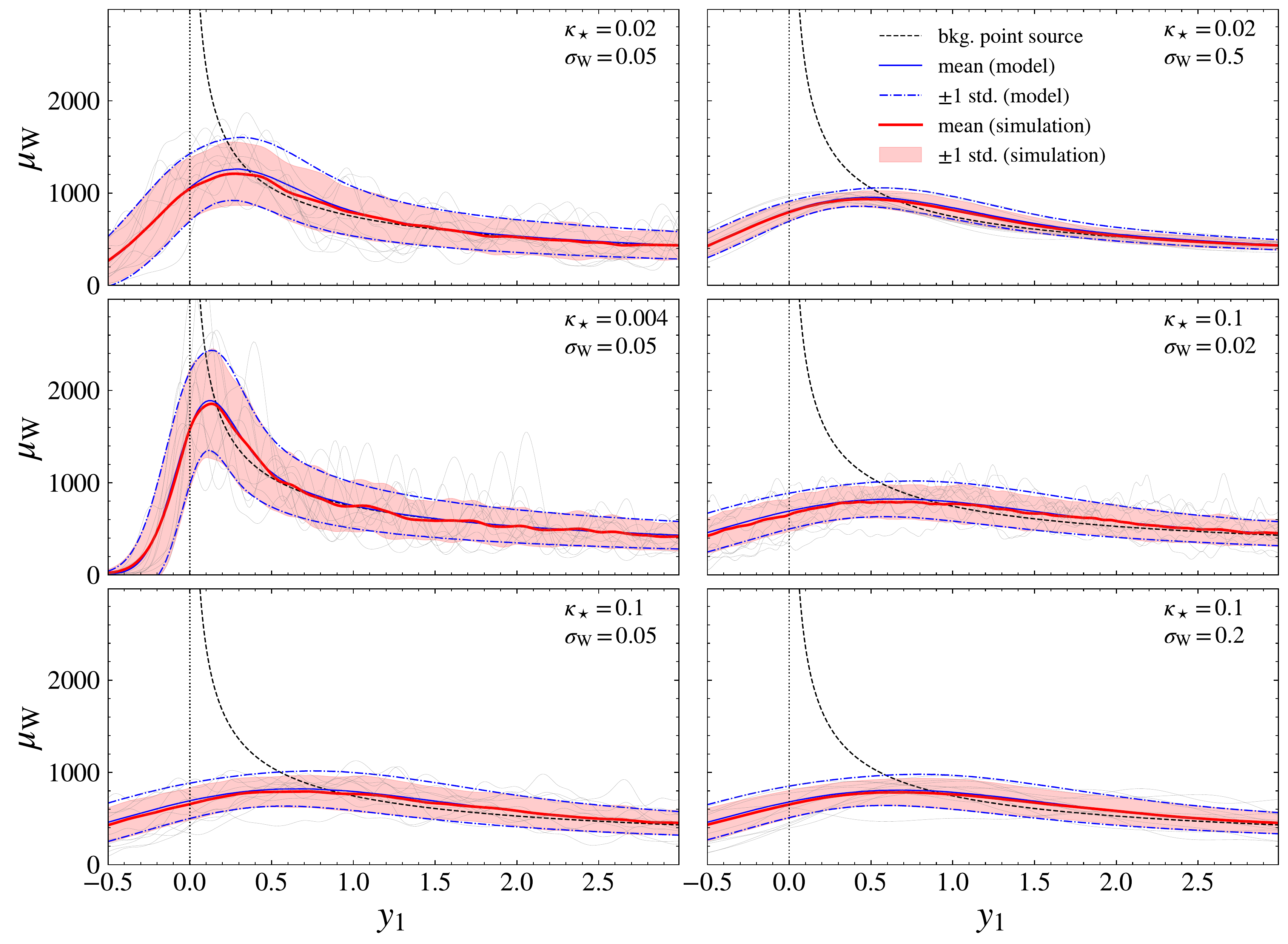}
    \caption{Statistics of the magnification factor $\mu_{\rm W}$ derived from numerical ray-shooting as a function of the distance $y_1$ to the macro fold caustic. We set parameter values $\theta_\star=1$, $d=10^{-5}$ and $R_*=1500$. Results are shown in separate panels for several choices of the microlens surface abundance $\kappa_\star$ and the source size $\sigma_{\rm W}$, all in the regime that the source overlaps multiple micro caustics. Theoretical calculations for the mean magnification $\VEV{\mu_{\rm W}}$ (solid blue) and its standard deviation ${\rm Std}[{\mu_{\rm W}}]$ (dash-dotted blue), all predicted by the semi-analytic model of this work, agree well with the numerical statistics (solid red curve for $\VEV{\mu_{\rm W}}$ and light red band for ${\rm Std}[{\mu_{\rm W}}]$) derived from 200 independent microlensing realizations. Additionally, 10 random realizations are shown as the grey curves. In all panels, the macro caustic is located at $y_1=0$ (vertical dotted line), and the magnification factor for a point source is shown as the dashed black curve.}
    \label{fig:num_check_muW}
\end{figure}

We numerically derive $\VEV{\mu_{\rm W}(\bfy)}$ and $\VEV{\mu_{\rm W}(\bfy)^2}$ by averaging over many random realizations for the microlenses. To be consistent with the averaging procedure we adopt in the numerical experiment, we always set $R_*=1500$ when evaluating $C^{\rm ml}(0)$, $C^{\rm ml}_\parallel(r)$ and $C^{\rm ml}_\perp(r)$; this is different from setting $R_*$ to be the clustering size of the micro images as proposed in \cite{Katz1986RandomScattering}. As we show in \reffig{num_check_muW}, for a range of parameters our semi-analytic calculations agree with numerical results to high accuracy, provided that the source typically overlaps with multiple micro caustics. The magnification $\mu_{\rm W}(\bfy)$ in fact can have a rather skewed, non-Gaussian distribution when the relative fluctuation is large, while the mean and variance are still accurately predicted by our semi-analytic formulae.

An interesting observation can be made from comparing the last three panels of \reffig{num_check_muW}: $\VEV{\mu_{\rm W}(\bfy)}$ and $\VEV{\mu_{\rm W}(\bfy)^2}$ in fact become insensitive to the source size $\sigma_{\rm W}$ if $\theta_\star\,\kappa^{1/2}_\star \gg \sigma_{\rm W}$, even though that the light curves are qualitatively distinct. The numerical results show that as $\sigma_{\rm W}$ decreases, the light curve becomes increasingly non-Gaussian while preserves $\VEV{\mu_{\rm W}(\bfy)}$ and $\VEV{\mu_{\rm W}(\bfy)^2}$! It is therefore reasonable to hypothesize that as long as the source size is much smaller than $\theta_\star\,\kappa^{1/2}_\star$, the results for $\VEV{\mu_{\rm W}(\bfy)}$ and $\VEV{\mu_{\rm W}(\bfy)^2}$ are insensitive to the source profile either; if the Gaussian source is replaced with a uniform-disk source, our results for $\VEV{\mu_{\rm W}(\bfy)}$ and $\VEV{\mu_{\rm W}(\bfy)^2}$ should remain correct.

\begin{figure}[t]
    \centering
    \includegraphics[scale=0.53]{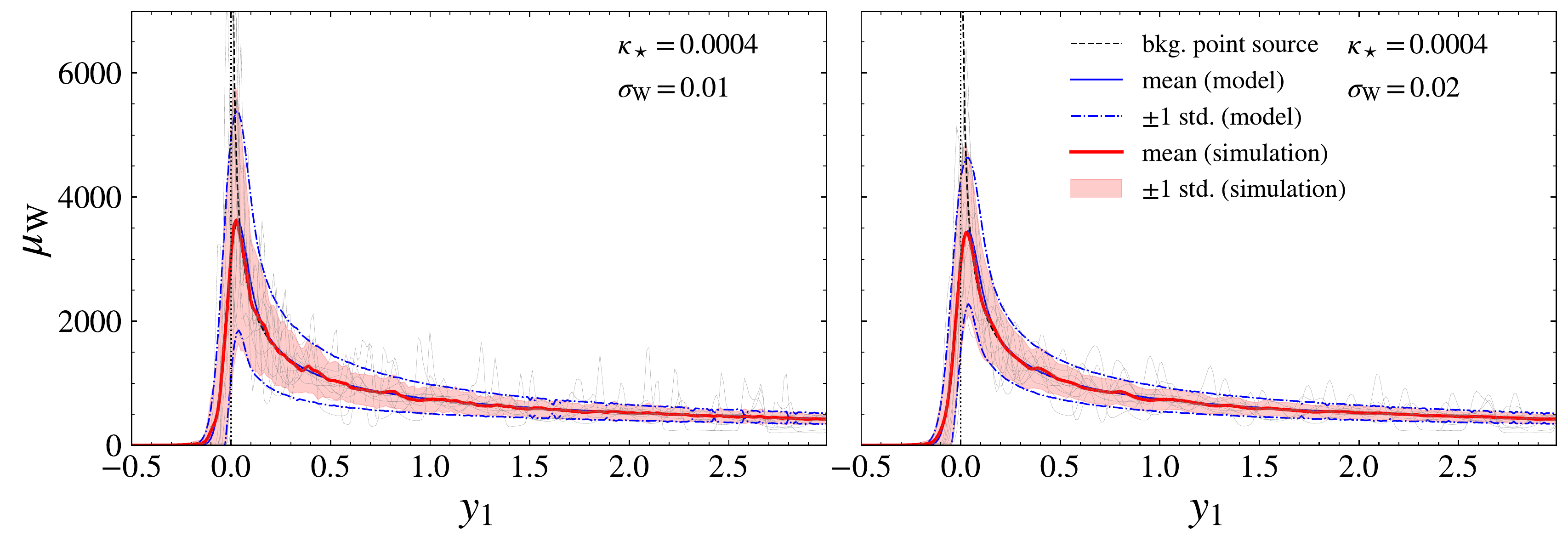}
    \caption{Same as \reffig{num_check_muW}, but for a relatively low microlens surface density $\kappa_\star=0.0004$ and reduced source sizes. Such small sources do not overlap multiple micro caustics, except in the very vicinity of the macro caustic where the macro magnification factor is sufficiently high $\mu_{\rm B} \gtrsim 2000$. In spite of the highly non-Gaussian nature of the magnification fluctuations, our semi-analytic predictions for the mean magnification and its variance still agree well with numerical simulations.}
    \label{fig:num_check_muW_2}
\end{figure}

The cases we examine in \reffig{num_check_muW} all correspond to a sufficiently large source size $\sigma_{\rm W}$ that overlaps multiple or at least order unity micro caustics, regardless of the size of the microlensing broadening $\theta_\star\,\kappa^{1/2}_\star$. As we can see from the ``light curves'', the fluctuations of the magnification factor are Gaussian or weakly non-Gaussian. To further test the range of validity of our approximation, in \reffig{num_check_muW_2} we examine cases where the number density of micro caustics are reduced and the source size $\sigma_{\rm W}$ is made smaller. The fluctuations of the magnification factor become highly non-Gaussian and very dramatic, approaching the familiar behavior of small sources exhibiting intermittent ``flares'' at micro caustic crossings. In these cases, the physical source extent hardly overlaps multiple micro caustics, while our semi-analytic approximation remains successful. We note that in these cases multiple micro images still arise (albeit the number of micro images is small), which may explain the success of the approximation. Hence, we find robust numerical evidences that the semi-analytic approximation developed in this work is applicable to computing the mean and variance of the magnification factor over a wide range of parameters.

\subsection{microlens mass distribution}
\label{sec:mlmassfun}

So far, results have been derived assuming identical microlens masses. Generalization to an arbitrary distribution of Einstein radii is straightforward if microlenses of different masses thoroughly mix in space. Introduce a differential contribution of the microlens convergence $\rmd \kappa_\star/\rmd \ln\theta^2_\star$. Formally, \refeq{sigma2ml} must be modified to
\begin{align}
\label{eq:sigml2massfun}
    \sigma^2_{\rm ml}(R_*,\,\ell_*) = \left( 1 - \gamma_E + \ln\frac{2\,R_*}{\overline{\theta^2_\star}\,\ell_*} \right)\,\left( \int\rmd\ln\theta^2_\star\,\frac{\rmd \kappa_\star}{\rmd \ln\,\theta^2_\star}\,\theta^2_\star \right) - \left( \int\rmd\ln\theta^2_\star\,\frac{\rmd \kappa_\star}{\rmd \ln\,\theta^2_\star}\theta^2_\star\,\ln\frac{\theta^2_\star}{\overline{\theta^2_\star}} \right),
\end{align}
where $\overline{\theta^2_\star}$ is the squared Einstein radius for the mean microlens mass. If we still set $\ell_*=1/\sigma_{\rm eff}$, then in the defining equation for $\sigma_{\rm eff}$, \refeq{sigmaeff}, $\kappa_\star\,\theta^2_\star$ also needs to be modified similarly to account for a distribution of Einstein radii. Following a similar logic, in using \refeq{lnPhi2} we must replace the first line with $-(1/2)\,(|\bfl_1|^2 + |\bfl_2|^2)$ multiplying \refeq{sigml2massfun}, and replace $\kappa_\star\,\theta^2_\star$ in the second line with the appropriate averaged quantity $\int\rmd\ln\theta^2_\star\,(\rmd \kappa_\star/\rmd \ln\,\theta^2_\star)\,\theta^2_\star$.

If microlenses do not differ in mass by orders of magnitude, the second integral is expected to be suppressed by the logarithmic factor, while the first integral is proportional to the average {\it squared} mass $\overline{\theta^4_\star}$ (i.e. weighted toward the more massive microlenses). However, the second integral may not be small at all when there is a hierarchy in $\theta^2_\star$. Interestingly, for galactic or intracluster stars a large mass hierarchy does exist between the sub-solar main-sequence (MS) dwarfs and the remnant black holes (BHs). For an old stellar population of which all stars with initial masses $> 1\,\Msun$ have become stellar remnants, and assuming an IMF $\rmd \phi(M)/\rmd M \propto M^{-2}$ for $M> 0.5\,\Msun$, about $\sim 0.007$ BH is expected for every MS dwarf~\citep{SatoshiTakada2021BHML}. Using a typical mass $0.3\,\Msun$ for the MSs and $8\,\Msun$ for the BHs, the BHs can make a comparable contribution to $\overline{\theta^4_\star}$, if not more, than the MSs. This implies the importance of BH microlenses in broadening the ``point spread function'' of random deflections despite their low number fraction. A detailed investigation into a realistic mass function will be included in a future work.

\section{Discussion}
\label{sec:discuss}

As we have explained, stochastic microlensing has a profound effect on the magnification factor of a lensed source. Now we discuss this more in the context of strong lensing produced by galaxy and galaxy cluster lenses.

The characteristic scale of random microlensing deflection for a microlens mass $M_\star$ corresponds to a source-plane scale
\begin{align}
    R_\star = 3\,\kappa^{1/2}_\star\,\theta_\star\,D_S \approx 2500\,{\rm AU}\,\left( \frac{\kappa_\star}{0.3} \right)^{1/2}\,\left( \frac{M_\star}{0.3\,\Msun} \right)^{1/2}\,\left( \frac{D_{LS}\,D_S\,D^{-1}_L}{{\rm Gpc}} \right)^{1/2},
\end{align}
where $D_L$, $D_S$ and $D_{LS}$ are the angular diameter distances to the lens plane, to the source plane, and from the lens plane to the source plane, respectively. We have multiplied by a factor of 3 to account for the Coulomb logarithm (see \refeq{sigma2ml}). In galactic lenses, the surface density of stellar microlenses is high $\kappa_\star \simeq 0.1$--$1$; hence $R_\star$ is smaller than the typical size of a star cluster, comparable to or larger than the sizes of optical quasars $\sim 10^3\,$AU~\citep{BlackBurne2011QSOmicrolensing}, while certainly larger than individual stellar photospheres. In galaxy cluster lenses, the intracluster stars have a substantially lower surface density $\kappa_\star \sim 0.001$--$0.01$, and $R_\star$ can be reduced by up to a factor of ten.

As long as the source's physical size is smaller than $R_\star$, the highest persistent magnification factor $\VEV{ \mu_{\rm W}(\bfy)}$ reached at a macro caustic is on the order of
\begin{align}
\label{eq:muWmeanmax}
    \VEV{\mu_{\rm W}}_{\rm max} \simeq \frac{\left(2\,d\,R_\star/D_S \right)^{-1/2}}{1-\kappa_0}\, \approx 400\,(1 - \kappa_0)^{-1}\,\left( \frac{d^{-1}}{1\arcsec} \right)^{1/2}\,\left( \frac{\kappa_\star}{0.3} \right)^{-1/4}\,\left( \frac{M_\star}{0.3\,\Msun} \right)^{-1/2}\,\left( \frac{D_L\,D_S\,D^{-1}_{LS}}{{\rm Gpc}} \right)^{1/2}.
\end{align}
For galaxy lenses, the choices $d^{-1} \sim 1\arcsec$ and $\kappa_\star \sim 0.1$ are reasonable, and hence only sufficiently compact sources can possibly acquire a {\it temporary} magnification much higher than $1000$, at micro caustic crossings. This still requires that $\mu_{\rm W}$ can fluctuate to a value much higher than $\VEV{\mu_{\rm W}(\bfy)}$. A conservative constraint is that the source size $\sigma_{\rm W}$ is smaller than $\sim \theta_\star\,\kappa^{-1/2}_\star/\VEV{\mu_{\rm W}}_{\rm max}$, the typical separation of micro caustics on the source plane. This limits the source size to
\begin{align}
\label{eq:sigWDSconstr}
    \sigma_{\rm W}\,D_S \lesssim 6\,{\rm AU}\,(1-\kappa_0)\,\left( \frac{d^{-1}}{1\arcsec} \right)^{-1/2}\,\left( \frac{\kappa_\star}{0.1} \right)^{-1/4}\,\left( \frac{M_\star}{0.3\,\Msun} \right)^{3/2}\,\left( \frac{D_L\,D_S\,D^{-1}_{LS}}{{\rm Gpc}} \right)^{-3/2}\,\left( \frac{D_S}{{\rm Gpc}} \right),
\end{align}
for which only individual source stars meet the requirement. However, the fluctuation in $\mu_{\rm W}$ can still be significantly suppressed for small sources if many micro images form ($\sigma_{\rm eff} > \theta_\star\,\kappa^{-1/2}_\star/\VEV{\mu_{\rm W}}_{\rm max}$), because only one pair of micro images are enhanced at each micro caustic crossing. 

\begin{figure}[t]
    \centering
    \includegraphics[scale=0.53]{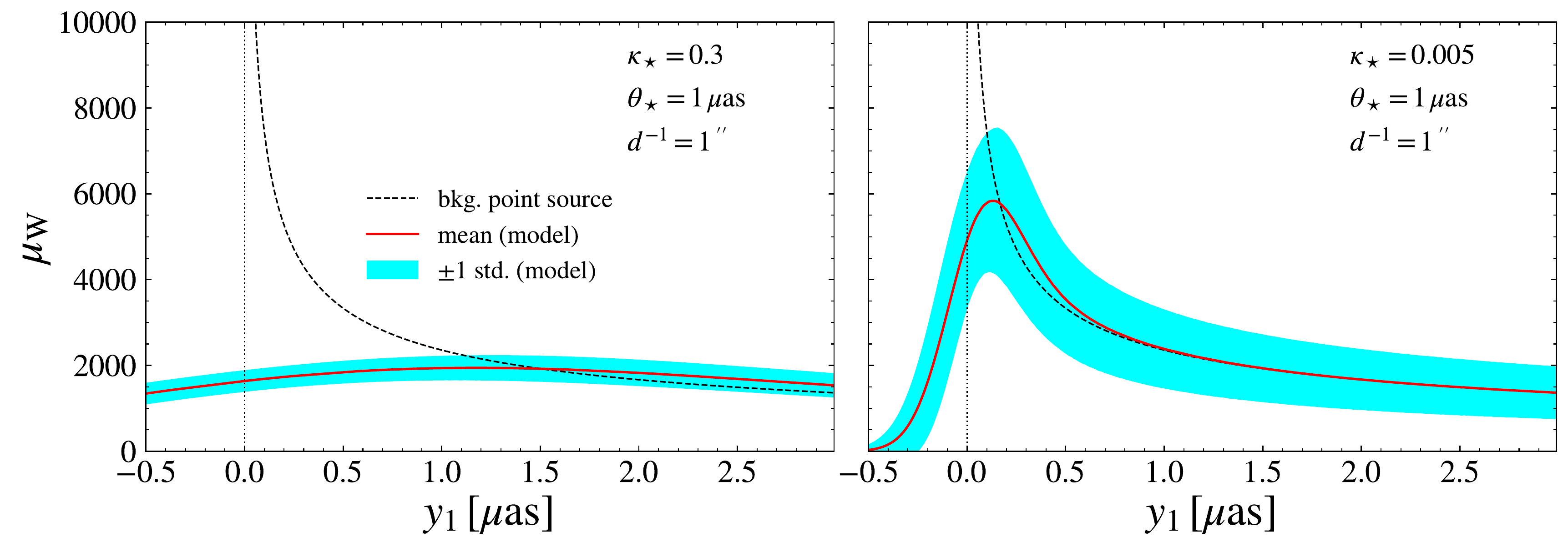}
    \caption{The mean and variance of the magnification factor $\mu_{\rm W}$ in the vicinity of a macro caustics with $\kappa_0=0.7$ and $d^{-1}=1\arcsec$ computed using the semi-analytic approximation developed in this work. We contrast between a high microlens surface density $\kappa_\star=0.3$ (left), typical of galaxy lensing, and a low microlens surface density $\kappa_\star=0.005$ (right), typical of cluster lensing. Both the mean magnification and the fluctuation around it are significantly suppressed in the former case by the excessively high number of micro caustics. While $\VEV{\mu_{\rm W}}$ and $\VEV{\mu^2_{\rm W}}$ are insensitive to the source size $\sigma_{\rm W}$ as long as $\sigma_{\rm W} \ll \theta_\star\,\kappa^{1/2}_\star$ (true for stellar photospheres), the magnification distribution becomes increasingly non-Gaussian for smaller sources.}
    \label{fig:muW_gal_vs_cluster}
\end{figure}

The best opportunities to have very high temporary magnifications for individual stars are to be found in galaxy cluster lensing with a small $\kappa_\star \sim 0.001$--$0.01$, for which the maximum values (at the tail of the distribution) can range from a few thousands to $10^4$~\citep{Diego2019ExtremeMagnificationUniverse}. The maximal mean magnification \refeq{muWmeanmax} can now reach $\VEV{\mu_{\rm W}}_{\rm max} \sim 3000\,(1-\kappa_0)^{-1}$ for $\kappa_\star=0.01$ and $d^{-1}=10\,\arcsec$, and the constraint on the source size \refeq{sigWDSconstr} is relaxed, which is also helped by the fact that the typical value of $d$ is reduced in cluster lenses. For galaxy lenses with $\kappa_\star \gtrsim 0.1$, we do not expect the magnification to strongly fluctuate and reach significantly higher than $1000$, as the corrugated micro caustic network is too dense. This conclusion is also reached from the argument that the peak magnification at a micro caustic crossing scales as $\kappa^{-3/4}_\star$~\citep{2017ApJ...850...49V}. From the perspective of this work, if we choose $\kappa_0=0.7$, $d^{-1}=1\arcsec$, $\theta_\star=1\,\mu{\rm as}$ and $\kappa_\star=0.3$, our semi-analytic approximation for $\VEV{\mu_{\rm W}}$ and $\VEV{\mu^2_{\rm W}}$ is applicable because $\sigma_{\rm eff} > \theta_\star\,\kappa^{-1/2}_\star/\VEV{\mu_{\rm W}}_{\rm max}$, and a small standard deviation $\sim 10$--$20 \%$ for the fractional magnification fluctuation around $\VEV{\mu_{\rm W}} \sim 1000$--$2000$ is predicted, insensitive to the source size. If we instead set $\kappa_\star=0.005$, for the same macro caustic, we find a larger standard deviation $\gtrsim 30 \%$ for the fractional magnification fluctuation, around a much higher mean $\VEV{\mu_{\rm W}} \sim 2000$--$6000$. This example is shown in \reffig{muW_gal_vs_cluster}.

It is worth to note that sub-galactic DM subhalos as substructure lenses tend to strongly perturb a galactic or cluster caustic and create secondary caustics under suitable conditions~\citep{2018ApJ...867...24D, Dai2020ArcSymmetryS1226}. An interesting consequence of these subhalos is then to increase the typical value of $d$ (i.e. weaken the caustic strength) and hence further reduce the allowed maximum mean magnification, in both galaxy and cluster lenses.

Applying the same analysis to a variety of large sources $\gtrsim$ tens of AUs, which include quasars, SNe~\citep{Kelly2015SNRefsdal, Goobar2017iPTF16geu}, or bloated stellar photospheres due to outburst or mass ejection, $\mu_{\rm W}$ is expected to fluctuate only mildly around the mean value in \refeq{muWmeanmax}, so magnifications significantly higher than $\sim 1000$ are prohibited by microlensing, even for galaxy cluster lenses. While multiply-imaged quasars commonly have magnification factors on the order $\mathcal{O}(10)$, quasars magnified by a hundred to a thousand fold are rarely reported. \cite{Fujimoto2020UltraluminousQuasar} suggested a candidate lensed quasar with a total magnification $\sim 450$. However, analysis of the proximity zone does not seem to support this idea~\citep{Davies2020LensedQSOproximity}. In another quadruply-imaged quasar, a ten-fold magnification anomaly was detected for one of the images, requiring a magnification factor as large as $\sim 100$~\citep{Glikman2018arXiv180705434G}. These large magnifications are likely to be consistent with the maximal values permitted by microlensing effects, if the caustic strength is not dramatically reduced by subhalos, i.e. $d^{-1} \gtrsim 0.1\arcsec$. We expect that microlensing effects dominate the truncation in the high magnification tail of the lensed quasars (especially for low-mass quasars and in the case of small Einstein radii), which may have implications for the impact of magnification bias on the luminosity function~\citep{PacucciLoeb2019LensedQSOzgt6, PacucciLoeb2020RealityMirage}.

\section{Conclusion}
\label{sec:concl}

Gravitationally lensed sources exhibit stochastic fluxes as a result of random microlensing if compact masses contribute a fraction of the lens surface mass. Through a first-principle statistical treatment of microlensing deflections, we have in this work derived a semi-analytic approximation for the mean and variance of the magnification factor, for a finite Gaussian source and for arbitrary macro lens models. A theoretically appealing feature of the new result is that the UV and IR logarithms are physically determined.

These general results are in the form of single and double image-plane integrals with simple and well-behaved integrands, and hence are practically useful as these can be efficiently evaluated using Monte Carlo integrators. Our analytic derivations suggest that the results are good approximations if the source of an effective size $\sigma_{\rm eff}$ overlaps multiple micro caustics, where the effective size is either the source's physical size $\sigma_{\rm W}$ or the scale of random microlensing deflections $\simeq \theta_\star\,\kappa^{1/2}_\star$, whichever is larger. Using numerical ray-shooting with random microlens realizations, we have demonstrated the accuracy of the approximation, even in cases where the microlensing-induced light curves are highly non-Gaussian. 

While we have specifically examined highly magnified sources near a macro fold caustic, for which a small convergence from the microlenses can induce dramatic flux variance, our results are readily applicable to other macro caustics, such as a cusp caustic, or higher-order catastrophes~\citep{Feldbrugge2019PichardLifshitz}. We have pointed out that the maximal magnification that can be realized at a macro fold caustic is not only limited by the source size $\sigma_{\rm W}$, but also by the characteristic scale of microlensing deflections $\simeq \theta_\star\,\kappa^{1/2}_\star$, especially for a source that overlaps multiple micro caustics or has many micro images.

Future work may adopt the formalism here to study the correlation of the magnification factor between two different source-plane positions, i.e. $\VEV{\mu_{\rm w}(\bfy_1)\,\mu_{\rm w}(\bfy_2)}$. For a moving source, this translates to the temporal correlation of microlensing lightcurves~\citep{WyitheTurner2002MicrolensingVariability, LewisIrwin1996MLIITemporalAnalysis, Neindorf2003extragalML}, and will be useful for interpreting cadence observations. Another interesting question regards the third-order moments of the magnification factor, as well as higher-order moments, which characterize the departure from Gaussian statistics. Our formalism may be applicable to the computation of these moments, which can help with the analyses of highly non-Gaussian light curves.

\begin{acknowledgments}
    
The authors thank Brenda Frye and Jordi Miralda-Escud\'{e} for inspiring discussions, and Jos\`{e} M. Diego for commenting on the draft paper near its completion. This research is supported under the startup grant provided as the Michael M. Garland Chair in Physics at the University of California, Berkeley. This material is based upon work supported by the National Science Foundation Graduate Research Fellowship under Grant No. DGE 1752814.

\end{acknowledgments}

\software{\texttt{Matplotlib} \citep{Matplotlib}, \texttt{vegas} \citep{vegasEnhanced}.}

\bibliographystyle{aasjournal}
\bibliography{cdmsubhalo,refs,refs2,refs3}

\appendix

\section{Integral for two-point deflection statistics}
\label{app:twoptinteg}

\begin{figure}[t]
    \centering
    \includegraphics[scale=0.3]{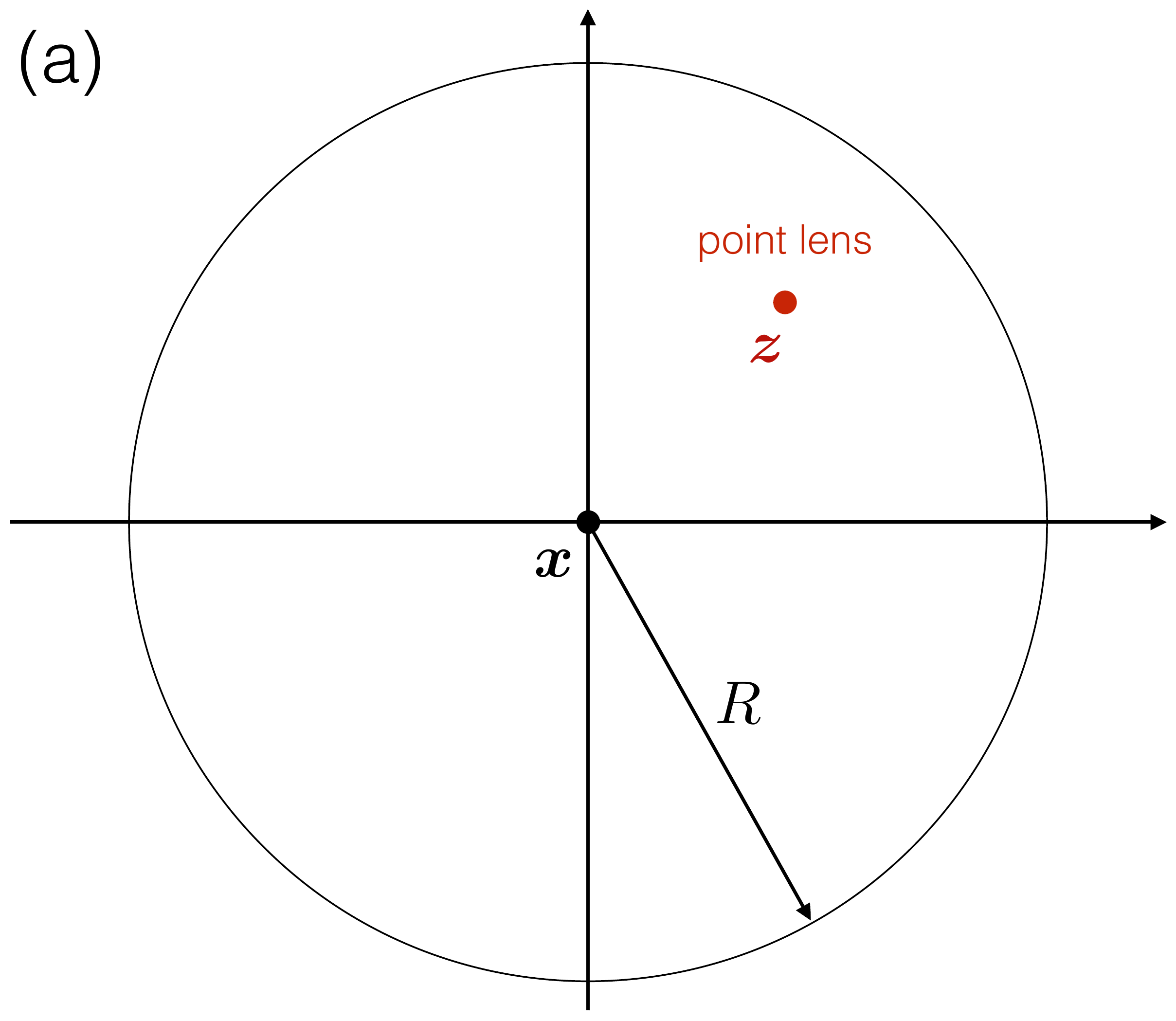}
    \hspace{0.5cm}
    \includegraphics[scale=0.3]{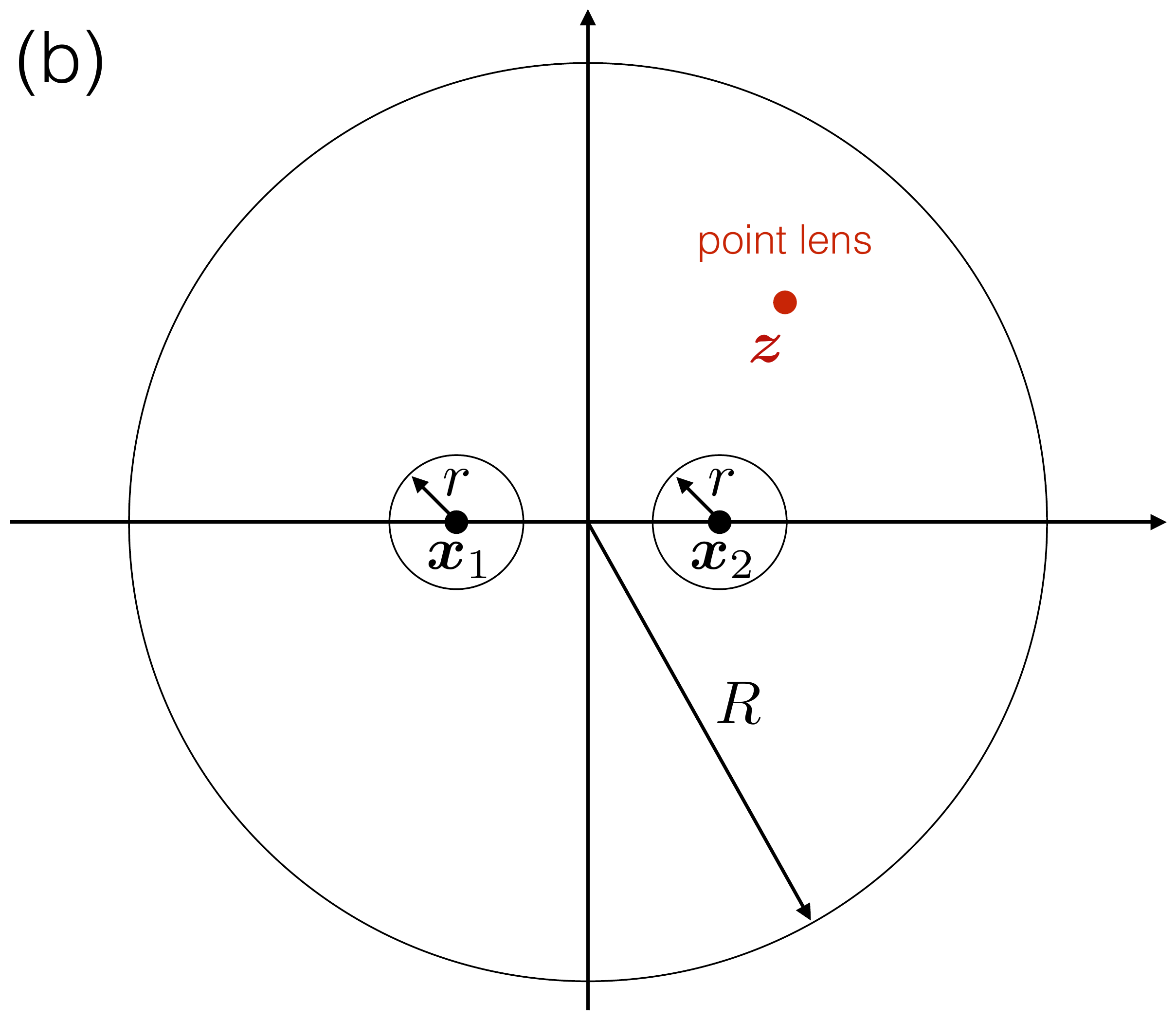}
    \caption{Integrals needed to find the characteristic functions for a single microlens: (a) one-point integral for $\varphi_1[\bfl;\,\bfx]$; (b) two-point integral for $\varphi_2[\bfl_1,\,\bfl_2;\,\bfx_1,\,\bfx_2]$.}
    \label{fig:ml_defl_integs}
\end{figure}

In this Appendix, we derive an approximation for the integral introduced in \refeq{lnvphi2}:
\begin{align}
\label{eq:2ptA1}
    \frac{1}{\pi\,R^2}\,\int\displaylimits_{|\bfz|<R}\,\rmd^2\bfz\,\left[ \exp\left( i\,\theta^2_\star\,\left( \frac{\bfl_1\cdot(\bfx_1 - \bfz)}{|\bfx_1 - \bfz|^2} + \frac{\bfl_2\cdot(\bfx_2 - \bfz)}{|\bfx_2 - \bfz|^2} \right) \right) - 1 \right],
\end{align}
We are interested in the regime that $|\bfl_1|$ and $|\bfl_2|$ are typically on the order of $\ell_*$ and that $\theta^2_\star\,\ell_*/|\bfx_1 - \bfx_2| \ll 1$. We introduce $\bfr_{12} = \bfx_2 - \bfx_1$ and $r_{12}=|\bfr_{12}|$.

Since each microlens has a uniform distribution on the lens plane, we consider the symmetric configuration that $\bfx_1$ and $\bfx_2$ lie on the first axis and are symmetric about the second axis.

Since by construction $\VEV{\bfalp_{\rm ml}(\bfx)} = 0$ for any $\bfx$, without calculation we anticipate that $\ln\phi_2[\bfl_1,\,\bfl_2;\,\bfx_1,\,\bfx_2] = i\,(\theta^2_\star/R^2)\,(\bfl_1\cdot\bfx_1 + \bfl_2\cdot\bfx_2) + \cdots$ at linear order in the Fourier wave vectors $\bfl_1$ and $\bfl_2$. This ensures that $\ln\Phi_2[\bfl_1,\,\bfl_2;\,\bfx_1,\,\bfx_2]$ has no linear order terms in the wave vectors. The goal is to derive contributions that are quadratic in the wave vectors.

Let us decompose the integration region, the entire disk $|\bfz| < R$, into two small disks, $|\bfz - \bfx_1| < r$ and $|\bfz - \bfx_2| < r$, plus the remaining region. Since typically $|\bfx_1 - \bfx_2|=r_{12} \gg \theta^2_\star\,\ell_*$, it is possible to set a value for $r$ such that the two small disks are non-overlapping yet $\theta^2_\star\,\ell_*/r \ll 1$.

First let us perform the integral in either of the small disks. Take the small disk $|\bfz - \bfx_1|<r$ for example. While $\exp(i\,\theta^2_\star\,\bfl_1\cdot(\bfx_1 - \bfz)/|\bfx_1 - \bfz|^2)$ cannot be treated perturbatively, we can do that for $\exp(i\,\theta^2_\star\,\bfl_2\cdot(\bfx_2 - \bfz)/|\bfx_2 - \bfz|^2)$,
\begin{align}
\label{eq:smalldiskexp}
    \exp\left( i\,\theta^2_\star\, \frac{\bfl_2\cdot(\bfx_2 - \bfz)}{|\bfx_2 - \bfz|^2} \right) \approx 1 + i\,\theta^2_\star\,\frac{\bfl_2\cdot(\bfx_2 - \bfz)}{|\bfx_2 - \bfz|^2} - \frac12\,\theta^4_\star\,\frac{[ \bfl_2\cdot(\bfx_2 - \bfz) ]^2}{|\bfx_2 - \bfz|^4} + \mathcal{O}\left(\left( \frac{\theta^2_\star\,\ell_*}{r_{12}} \right)^3 \right),
\end{align}
because $\theta^2_\star\,\ell_*/r_{12} \ll 1$. The constant term in \refeq{smalldiskexp} corresponds to
\begin{eqnarray}
\label{eq:ConstQuad}
    \frac{1}{\pi\,R^2}\,\int\displaylimits_{|\bfz - \bfx_1|<r}\,\rmd^2\bfz\,\left[ \exp\left( i\,\theta^2_\star\, \frac{\bfl_1\cdot(\bfx_1 - \bfz)}{|\bfx_1 - \bfz|^2} \right) - 1 \right] & = & \frac{1}{\pi\,R^2}\,\int\displaylimits_{|\bfz|<r}\,\rmd^2\bfz\,\left( e^{ -i\,\theta^2_\star\, \bfl_1\cdot\bfz/|\bfz|^2} - 1 \right) \nonumber\\
    & \approx & - \frac{\theta^4_\star\,\ell^2_1}{2\,R^2}\,\left( 1 - \gamma_E + \ln\frac{2\,r}{\theta^2_\star\,\ell_*} \right).
\end{eqnarray}
The last step, referring to \refeq{Itinteg}, is justified because $\theta^2_\star\,\ell_1/r \ll 1$ for our choice of $r$. We have also set $\ell_1 \simeq \ell_*$ in the logarithm. The linear term in \refeq{smalldiskexp} makes a contribution at the quadratic order in the wave vectors:
\begin{eqnarray}
\label{eq:LinQuad}
    && \frac{1}{\pi\,R^2}\,\int\displaylimits_{|\bfz - \bfx_1|<r}\,\rmd^2\bfz\,\left( i\,\theta^2_\star\, \frac{\bfl_1\cdot(\bfx_1 - \bfz)}{|\bfx_1 - \bfz|^2} \right)\,\left( i\,\theta^2_\star\, \frac{\bfl_2\cdot(\bfx_2 - \bfz)}{|\bfx_2 - \bfz|^2} \right) = - \frac{\theta^4_\star}{\pi\,R^2}\,\int\displaylimits_{|\bfz|<r}\,\rmd^2\bfz\,\frac{\bfl_1\cdot\bfz}{|\bfz|^2}\,\frac{\bfl_2\cdot(\bfz - \bfr_{12})}{|\bfz - \bfr_{12}|^2} \nonumber\\
    & = & - \frac{\theta^4_\star\,r^2}{2\,R^2}\,\frac{1}{r^4_{12}}\,\left[ \left(\bfl_1\cdot\bfr_{12} \right)\,\left(\bfl_2\cdot\bfr_{12} \right) - \left(\bfl_1\times\bfr_{12} \right)\,\left(\bfl_2\times\bfr_{12} \right) \right].
\end{eqnarray}
Here we use the notation of vector cross product $\bfa \times \bfb := \epsilon_{ij}\,a_i\,b_j$, where $\epsilon_{ij}$ is the anti-symmetric tensor in two dimensions. The quadratic term in \refeq{smalldiskexp} makes a contribution at the quadratic order in the wave vectors:
\begin{eqnarray}
\label{eq:QuadQuad}
    && \frac{1}{\pi\,R^2}\,\int\displaylimits_{|\bfz - \bfx_1|<r}\,\rmd^2\bfz\,\left( - \frac12\,\theta^4_\star\,\frac{[ \bfl_2\cdot(\bfx_2 - \bfz) ]^2}{|\bfx_2 - \bfz|^4} \right) = - \frac{\theta^4_\star}{2}\,\frac{1}{\pi\,R^2}\,\int\displaylimits_{|\bfz|<r}\,\rmd^2\bfz\,\frac{[ \bfl_2\cdot(\bfz - \bfr_{12}) ]^2}{|\bfz - \bfr_{12}|^4} \nonumber\\
    & = & \frac{\theta^4_\star}{4\,R^2}\,\left[ \ln\left(  1 - \frac{r^2}{r^2_{12}} \right)\,|\bfl_2|^2 - \frac{r^2}{r^4_{12}}\,\left( \left( \bfl_2\cdot\bfr_{12} \right)^2 - \left( \bfl_2\times\bfr_{12} \right)^2 \right) \right].
\end{eqnarray}
Since we can choose $r$ as small as $\theta^2_\star\,\ell_*$ which is assumed to be much smaller than $r_{12}=|\bfr_{12}|$, \refeq{LinQuad} and \refeq{QuadQuad} are parametrically smaller than \refeq{ConstQuad} by a factor $(\theta^2_\star\,\ell_*/r_{12})^2 \ll 1$. The same analysis is applicable to the integral within the small disk $|\bfz - \bfx_2| < r$, yielding results under cross symmetry $\bfl_1 \leftrightarrow \bfl_2$ and $\bfx_1 \leftrightarrow \bfx_2$.

What remains is the integration over the large disk $|\bfz| < R$ but {\it excluding} the two small disks $|\bfz - \bfx_1| < r$ and $|\bfz - \bfx_2| < r$. In this region, it is justified to expand the exponent in \refeq{2ptA1}:
\begin{eqnarray}
    && \exp\left( i\,\theta^2_\star\,\left( \frac{\bfl_1\cdot(\bfx_1 - \bfz)}{|\bfx_1 - \bfz|^2} + \frac{\bfl_2\cdot(\bfx_2 - \bfz)}{|\bfx_2 - \bfz|^2} \right) \right) - 1 = i\,\theta^2_\star\,\left( \frac{\bfl_1\cdot(\bfx_1 - \bfz)}{|\bfx_1 - \bfz|^2} + \frac{\bfl_2\cdot(\bfx_2 - \bfz)}{|\bfx_2 - \bfz|^2} \right) \nonumber\\
    && - \frac12\,\theta^4_\star\,\left( \frac{[\bfl_1\cdot(\bfx_1 - \bfz)]^2}{|\bfx_1 - \bfz|^4} + \frac{[\bfl_2\cdot(\bfx_2 - \bfz)]^2}{|\bfx_2 - \bfz|^4} + 2\, \frac{\bfl_1\cdot(\bfx_1 - \bfz)}{|\bfx_1 - \bfz|^2}\,\frac{\bfl_2\cdot(\bfx_2 - \bfz)}{|\bfx_2 - \bfz|^2} \right) + \cdots
\end{eqnarray}
We only concern terms that have quadratic dependence on the Fourier wave vectors (the second line).

First, let us examine the contribution:
\begin{align}
    - \frac{\theta^4_\star}{2}\,\frac{1}{\pi\,R^2}\,\int\,\rmd^2\bfz\,\frac{[ \bfl_1\cdot(\bfx_1 - \bfz) ]^2}{|\bfx_1 - \bfz|^4}.
\end{align}
We are supposed to perform the integration over $|\bfz| < R$ excluding $|\bfz - \bfx_1| < r$ and $|\bfz - \bfx_2| < r$. However, in the regime $R \gg r$ and $|\bfr_{12}|\gg r$, we introduce an error suppressed by $(r/R)^2$ if instead we evaluate
\begin{align}
\label{eq:selfQuad}
    - \frac{\theta^4_\star}{2}\,\frac{1}{\pi\,R^2}\,\int\displaylimits_{r< |\bfz| < R}\,\rmd^2\bfz\,\frac{\left( \bfl_1\cdot \bfz \right)^2}{|\bfz|^4} = - \frac{\theta^4_\star\,\ell^2_1}{2\,R^2}\,\ln\frac{R}{r}.
\end{align}
Here, we essentially neglect the $|\bfz - \bfx_2|<r$ disk, and move the $|\bfz - \bfx_1|<r$ disk to the center of the $|\bfz|<R$ disk. Similarly, we have another contribution $-\theta^4_\star\,\ell^2_2/(2\,R^2)\,\ln(R/r)$ from the $\bfl_1 \leftrightarrow \bfl_2$ and $\bfx_1 \leftrightarrow \bfx_2$ crossed term.

What remains is the contribution
\begin{align}
    - \frac{\theta^4_\star}{\pi\,R^2}\,\int\,\rmd^2\bfz\,\frac{\bfl_1\cdot(\bfx_1 - \bfz)}{|\bfx_1 - \bfz|^2}\,\frac{\bfl_2\cdot(\bfx_2 - \bfz)}{|\bfx_2 - \bfz|^2}.
\end{align}
It is justified to modify the integration region to be just $|\bfz|<R$; we in fact do not have to subtract the two small disks at all, as doing that only account for negligibly small contributions suppressed by $(r/R)^2$. We find
\begin{eqnarray}
\label{eq:crossQaud}
    && - \frac{\theta^4_\star}{\pi\,R^2}\,\int\displaylimits_{|\bfz|<R}\,\rmd^2\bfz\,\frac{\bfl_1\cdot(\bfx_1 - \bfz)}{|\bfx_1 - \bfz|^2}\,\frac{\bfl_2\cdot(\bfx_2 - \bfz)}{|\bfx_2 - \bfz|^2} \nonumber\\
    & = & - \frac{\theta^4_\star}{R^2}\,\left[ \left( \ln\frac{2\,R}{r_{12}} + \frac12\,\ln\left( 1 + \frac{r^2_{12}}{4\,R^2} \right) - \ln 2 + \frac12 \right)\,\left( \bfl_1\cdot\bfl_2 \right) - \frac{\left(\bfl_1\cdot\bfr_{12}\right)\,\left(\bfl_2\cdot\bfr_{12}\right)}{r^2_{12}}\right].
\end{eqnarray}

After putting together \refeq{ConstQuad}, \refeq{selfQuad} and \refeq{crossQaud}, as well as their crossed symmetric counterpart terms under $\bfl_1 \leftrightarrow \bfl_2$ and $\bfx_1 \leftrightarrow \bfx_2$, we find an approximation:
\begin{eqnarray}
\label{eq:lnvphi2approx}
    \ln\varphi_2[\bfl_1,\,\bfl_2;\,\bfx_1,\,\bfx_2] & \approx & - \frac{\theta^4_\star}{2\,R^2}\,\left( |\bfl_1|^2 + |\bfl_2|^2 \right)\,\left( 1 - \gamma_E + \ln\frac{2\,R}{\theta^2_\star\,\ell_*} \right) \\
    && - \frac{\theta^4_\star}{R^2}\,\left[ \left( \ln\frac{2\,R}{r_{12}} + \frac12\,\ln\left( 1 + \frac{r^2_{12}}{4\,R^2} \right) - \ln 2 + \frac12 \right)\,\left( \bfl_1\cdot\bfl_2 \right) - \frac{\left(\bfl_1\cdot\bfr_{12}\right)\,\left(\bfl_2\cdot\bfr_{12}\right)}{r^2_{12}}\right].  \nonumber
\end{eqnarray}
As we might have anticipated, the radius of the small disks $r$ drops out of this leading result of quadratic dependence in the Fourier wave vectors. Despite having made various non-trivial approximations, the bottom line is that we have extracted all terms that have a quadratic dependence on $\bfell_1$ and $\bfell_2$ and survive the $r \rightarrow 0$ limit.

\section{Regularization of deflection correlation functions}
\label{app:Creg}

According to \refeq{CL} and \refeq{CT}, the longitudinal and transverse two-point correlation functions for the microlensing deflection, $C^{\rm ml}_\parallel(r)$ and $C^{\rm ml}_\perp(r)$, respectively, have a logarithmic divergence $-\kappa_\star\,\theta^2_\star\,\ln\,r$ at small separations $r \ll \theta^2_\star\,\ell_*$. This renders the covaiance matrix $\textbf{C}_{\rm ml}(\bfr)$ in \refeq{covCml} not positive definite, and hence the approximate expression for $\VEV{\mu_{\rm W}(\bfy)^2}$ (\refeq{muW2dblintegML}) invalid to evaluate numerically. At large separations $r\gg R_*$, on the other hand, the derivation of \refeq{lnPhi2} is unjustified. Nevertheless, the intuition is that contributions to the integral from these two problematic regimes are unimportant anyway, which implies a freedom to regularize the logarithmic divergences in these two regimes without invalidating the key results.

\begin{figure}[t]
    \centering
    \includegraphics[scale=0.7]{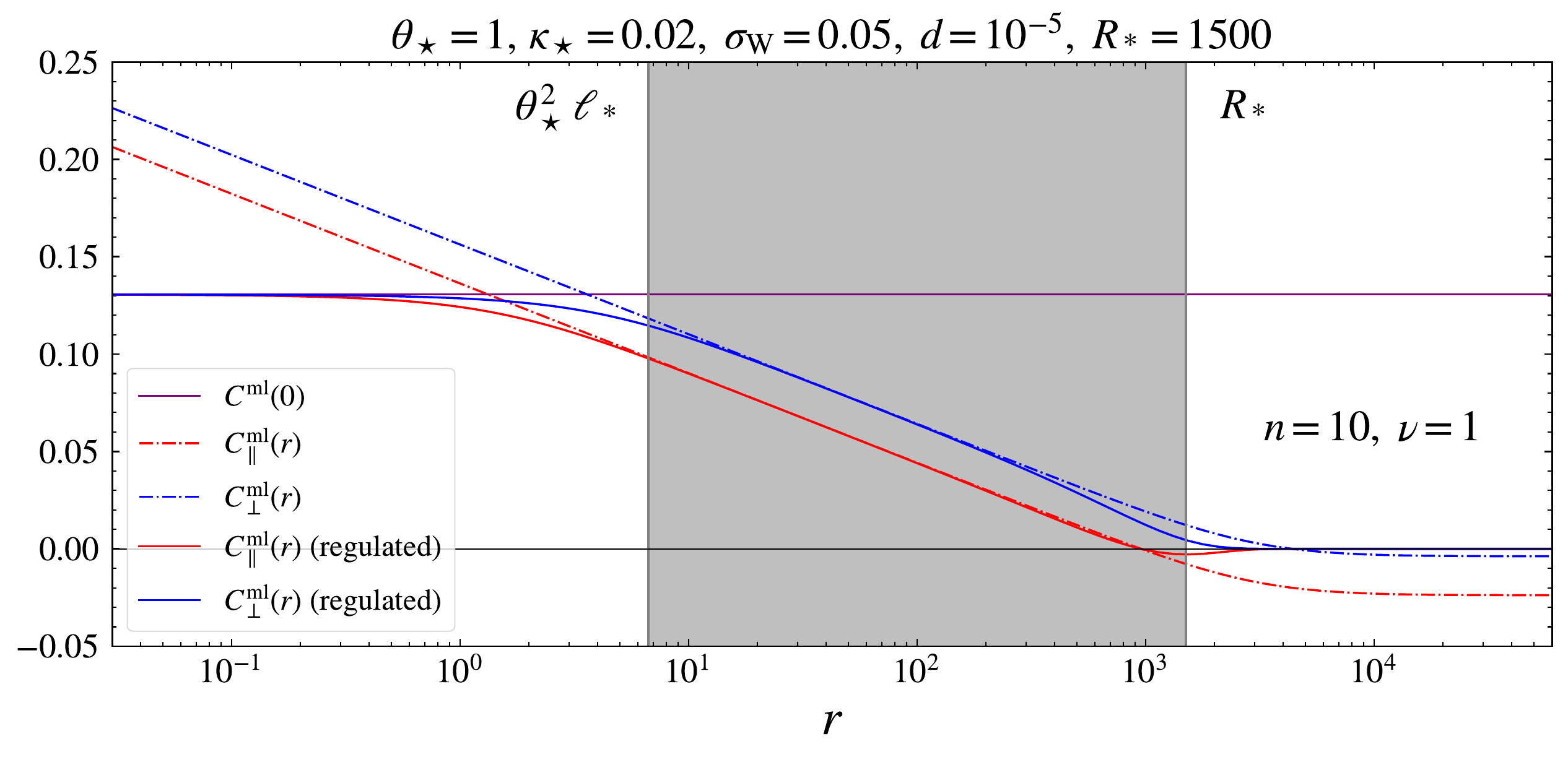}
    \caption{Unregulated and regulated two-point correlation functions $C^{\rm ml}_\parallel(r)$ and $C^{\rm ml}_\perp(r)$ compared to $C^{\rm ml}(0)$. The regularization procedure ensures that $C^{\rm ml}_\parallel(r)$ and $C^{\rm ml}_\perp(r)$ are no greater than $C^{\rm ml}(0)$ without modifying their simple logarithmic scaling behaviors in the key regime $\theta^2_\star\,\ell_* < r < R_*$ (region shaded in grey), which has the dominant contribution to the variance of $\mu_{\rm W}(\bfy)$.}
    \label{fig:Craa_reg}
\end{figure}

We seek a regularization scheme such that: (1) for $r \ll \theta^2_\star\,\ell_*$, $C^{\rm ml}_\parallel(r)$ and $C^{\rm ml}_\perp(r)$ are always smaller than but approach $C^{\rm ml}(0)$; (2) for $r \gg R_*$, $C^{\rm ml}_\parallel(r)$ and $C^{\rm ml}_\perp(r)$ are asymptotically zero. Specifically, we define the following regularized functions:
\begin{eqnarray}
    C^{\rm ml}_{X, {\rm reg}}(r) := C^{\rm ml}(0)\,\left[ 1 + \left( \frac{C^{\rm ml}(0)}{|C^{\rm ml}_X(r)|\,e^{-(r/(\nu\,R_*))^2}} \right)^n \right]^{-1/n}\,{\rm sgn}[C^{\rm ml}_X(r)],
\end{eqnarray}
where $X$ is either $\parallel$ or $\perp$. We find that a suitable choice for the two parameters are $n=10$ and $\nu = 1$. An numerical example is plotted in \reffig{Craa_reg}. As far as the computation of $\VEV{\mu_{\rm W}(\bfy)^2}$ is concerned, what is important is the regime $\theta^2_\star\,\ell_* < r < R_*$, where both $C^{\rm ml}_\parallel(r)$ and $C^{\rm ml}_\perp(r)$ are linear functions of $\ln r$.

\section{Uniform background convergence and shear}
\label{app:uniformbkg}

In this Appendix, we apply the analytic approximation derived in \refsec{microlens} to a special case: the background lens has a constant convergence $\kappa_{\rm B}$ and shear $(\gamma_{{\rm B}, 1},\,\gamma_{{\rm B}, 2})$. We can write
\begin{align}
    \bfx - \bfalp_{\rm B}(\bfx) = \left[ \begin{array}{cc}
    1 - \kappa_{\rm B} - \gamma_{{\rm B}, 1} & - \gamma_{{\rm B}, 2} \\
    - \gamma_{{\rm B}, 2} & 1 - \kappa_{\rm B} + \gamma_{{\rm B}, 1} \\
    \end{array} \right] \cdot \bfx \equiv \textbf{A}_{\rm B} \cdot \bfx,
\end{align}
where we have defined the background deformation matrix $\textbf{A}_{\rm B}$. Thus, the background lens model has translational invariance. Without loss of generality, we may set $\gamma_{{\rm B}, 1} = \gamma_{\rm B}$ and $\gamma_{{\rm B}, 2} = 0$. The moments of $\mu_{\rm W}(\bfy)$ are independent of $\bfy$, so we are free to set $\bfy = 0$.

\refeq{muWvevML} is trivially evaluated to
\begin{align}
    \VEV{\mu_{\rm W}} = \left| \left( 1 - \kappa_{\rm B} - \gamma_{\rm B} \right)\,\left( 1 - \kappa_{\rm B} + \gamma_{\rm B} \right) \right|^{-1}.
\end{align}
This equals the (uniform) background magnification factor, and is independent of $\sigma_{\rm ml}$ and $\sigma_{\rm W}$.

To evaluate $\VEV{\mu^2_{\rm W}}$ (see \refeq{muW2dblinteg}), we first rescale the image-plane variables $\tilde x_{1,1} = |1-\kappa_{\rm B} - \gamma_{\rm B}|\,x_{1,1}$, $\tilde x_{1,2} = |1-\kappa_{\rm B} + \gamma_{\rm B}|\,x_{1,2}$, $\tilde x_{2,1} = |1-\kappa_{\rm B} - \gamma_{\rm B}|\,x_{2,1}$, and $\tilde x_{2,2} = |1-\kappa_{\rm B} + \gamma_{\rm B}|\,x_{2,2}$. This generates a Jacobian
\begin{align}
    \rmd^2\bfx_1\,\rmd^2\bfx_2 =  \frac{\rmd^2\tilde\bfx_1\,\rmd^2\tilde\bfx_2}{\left| \left( 1 - \kappa_{\rm B} - \gamma_{\rm B} \right)\,\left( 1 - \kappa_{\rm B} + \gamma_{\rm B} \right) \right|^2}.
\end{align}
The double integration over $\tilde\bfx_1$ and $\tilde\bfx_2$ can be recast into an integration over the relative vector $\tilde\bfr := \tilde\bfx_2 - \tilde\bfx_1$ and the ``center of mass'' vector $\tilde\bfR := (\tilde\bfx_1 + \tilde\bfx_2)/2$ through a transformation of variables $\rmd^2\tilde\bfx_1\,\rmd^2\tilde\bfx_2 = \rmd^2\tilde\bfR\,\rmd^2\tilde\bfr$. Since the covariance matrix $\textbf{C}(\bfr)$ only depends on $\tilde\bfr$ but not on $\tilde\bfR$, the integration over $\tilde\bfR$ is a two-dimensional Gaussian integral and can be analytically evaluated, leaving behind the $\tilde\bfr$ integral:
\begin{align}
\label{eq:muW2cstbkg}
    \VEV{\mu^2_{\rm W}} = \left(4\pi\right)^{-1}\,\left| \left( 1 - \kappa_{\rm B} - \gamma_{\rm B} \right)\,\left( 1 - \kappa_{\rm B} + \gamma_{\rm B} \right) \right|^{-2}\,\int\,\rmd^2\tilde\bfr\,\frac{\exp\left(- \frac12\,\tilde\bfr^T\cdot\textbf{D}(\bfr)\cdot \tilde\bfr \right)}{\sqrt{( C^{\rm ml}(0) + \sigma^2_{\rm W} - C^{\rm ml}_\parallel(r) )\,( C^{\rm ml}(0) + \sigma^2_{\rm W} - C^{\rm ml}_\perp(r) ) }},
\end{align}
where the two-by-two symmetric matrix $\textbf{D}(\bfr)$ have matrix elements:
\begin{eqnarray}
    D_{11} & = & \frac{2\,C^{\rm ml}(0) + 2\,\sigma^2_{\rm W} - \left( C^{\rm ml}_\parallel(r) + C^{\rm ml}_\perp(r) \right) + \left( C^{\rm ml}_\parallel(r) - C^{\rm ml}_\perp(r) \right)\,\cos 2\varphi}{4\,\left(C^{\rm ml}(0) + \sigma^2_{\rm W} - C^{\rm ml}_\parallel(r) \right)\,\left( C^{\rm ml}(0) + \sigma^2_{\rm W} - C^{\rm ml}_\perp(r)\right) }, \\
    D_{12} = D_{21} & = & \frac{\left( C^{\rm ml}_\parallel(r) - C^{\rm ml}_\perp(r) \right)\,\sin 2\varphi}{4\,\left(C^{\rm ml}(0) + \sigma^2_{\rm W} - C^{\rm ml}_\parallel(r) \right)\,\left( C^{\rm ml}(0) + \sigma^2_{\rm W} - C^{\rm ml}_\perp(r)\right)}, \\
    D_{22} & = & \frac{2\,C^{\rm ml}(0) + 2\,\sigma^2_{\rm W} - \left( C^{\rm ml}_\parallel(r) + C^{\rm ml}_\perp(r) \right) - \left( C^{\rm ml}_\parallel(r) - C^{\rm ml}_\perp(r) \right)\,\cos 2\varphi}{4\,\left(C^{\rm ml}(0) + \sigma^2_{\rm W} - C^{\rm ml}_\parallel(r) \right)\,\left( C^{\rm ml}(0) + \sigma^2_{\rm W} - C^{\rm ml}_\perp(r)\right)}.
\end{eqnarray}
The integral \refeq{muW2cstbkg} can be numerically evaluated given $\kappa_{\rm B}$ and $\gamma_{\rm B}$. 

The result can be further simplified if the background shear vanishes $\gamma_{\rm B}=0$. In this case, $\tilde\bfr = |1-\kappa_{\rm B}|\,\bfr$. We derive
\begin{align}
\label{eq:muW2noshear}
    \VEV{\mu^2_{\rm W}} = \frac{1}{2\,(1-
\kappa_{\rm B})^2}\,\int^\infty_0\,\frac{ r\,\rmd r\,e^{- \frac{(1-\kappa_{\rm B})^2\,r^2}{4\,\left( C^{\rm ml}(0) + \sigma^2_{\rm W} - C^{\rm ml}_\parallel(r) \right)}}}{\sqrt{( C^{\rm ml}(0) + \sigma^2_{\rm W} - C^{\rm ml}_\parallel(r) )\,( C^{\rm ml}(0) + \sigma^2_{\rm W} - C^{\rm ml}_\perp(r) ) }}.
\end{align}
Previously, \cite{RefsdalStabell1991MLLargeSources} (also see \cite{Seitz1994microlensingII}) showed that in the case $\kappa_{\rm B}=\kappa_\star \ll 1$ and $\gamma_{\rm B}=0$, and for a large disk source of (source-plane) angular radius $\theta_{\rm S}$ and uniform surface brightness, the variance in the relative magnification factor is given by
\begin{align}
    \frac{\VEV{\mu^2_{\rm W}}- \VEV{\mu_{\rm W}}^2}{\VEV{\mu_{\rm W}}^2} \approx 4\,\kappa_\star\,\frac{\theta^2_\star}{\theta^2_{\rm S}}, \quad\quad ({\rm uniform\,\,disk\,\,source}).
\end{align}
The result for a large source of a two-dimensional Gaussian surface brightness profile must be modified. Following the reasoning of \cite{RefsdalStabell1991MLLargeSources}, we derive that the correct formula has a different coefficient:
\begin{align}
\label{eq:2dgaussianlimit}
    \frac{\VEV{\mu^2_{\rm W}}- \VEV{\mu_{\rm W}}^2}{\VEV{\mu_{\rm W}}^2} \approx \kappa_\star\,\frac{\theta^2_\star}{\sigma^2_{\rm W}}, \quad\quad ({\rm 2D\,\,Gaussian\,\,source}),
\end{align}
which agrees with \cite{Neindorf2003extragalML}. This limiting case can be explicitly verified by Taylor expanding the integrand of \refeq{muW2noshear} to the next-to-leading order in $1/\sigma^2_{\rm W}$ and then integrating term by term. We need to account for the logarithmic terms $\sim \ln r$ in \refeq{CL} and \refeq{CT} (otherwise an incorrect coefficient of $1/2$ would be obtained!), and the fact that $C^{\rm ml}_\perp(r) - C^{\rm ml}_\parallel(r) = 1$. Hence, our results reproduce the limiting case of \refeq{2dgaussianlimit}.

\cite{DeguchiWatson1987muvariance} computed $\VEV{\mu^2_{\rm W}}/\VEV{\mu_{\rm W}}^2$ for the cases $\gamma_{\rm B}=0$ and arbitrary $\kappa_{\rm B} = \kappa_\star > 0$, for different source sizes (see Figure 1 therein). Our \refeq{muW2cstbkg} is applicable, and hence can be examined for a comparison, if the source size is larger than the typical separation of micro caustics, i.e. $\sigma_{\rm W} \gtrsim \theta_\star\,\kappa_\star^{-1/2}$ ($a/a_0 \gtrsim \tau^{-1/2}$ in their notation). Good numerical agreement is found in that regime. We also find good numerical agreement with Figure 1 of \cite{Seitz1994microlensingII}. Motivated by the study of microlensing of extragalactic stars in lensed galaxies, \cite{Tuntsov2004compactDMFromClusters} presented a different semi-analytic calculations for the magnification variance in the presence of background shear $\gamma_{\rm B}\neq 0$. However, they presented reliable results mainly in the regime of small source sizes, and they also pointed out that their numerical results might not be accurate when the background magnification factor is high, i.e. $(1-\kappa_*)^2 - \gamma^2_{\rm B} \approx 0$.



\end{document}